\documentclass[fleqn,usenatbib]{mnras}

\usepackage{newtxtext,newtxmath}
\usepackage[T1]{fontenc}

\DeclareRobustCommand{\VAN}[3]{#2}
\let\VANthebibliography\thebibliography
\def\thebibliography{\DeclareRobustCommand{\VAN}[3]{##3}\VANthebibliography}

\usepackage{graphicx}	
\usepackage{amsmath}	
\usepackage{color}

\newcommand{\rp}{r_+}
\newcommand{\der}{\mathrm{d}}
\newcommand{\vel}{\vartheta}
\newcommand{\rs}{r_{\rm s}}
\newcommand{\mbh}{M_{\rm BH}}
\newcommand{\ie}{\it i.e.}
\newcommand{\mdot}{\dot {M}}
\newcommand{\be}{\cal E}
\newcommand{\entmdot}{{\mathcal {\dot{M}}}}
\newcommand{\denc}{\mathcal{D}_\mathrm{c}}
\newcommand{\numc}{\mathcal{N}_\mathrm{c}}

\newcommand{\rc}{r_\mathrm{c}}

\title[Spin effect on accretion flows]{Effect of spin on the dynamics of multi-component trans-relativistic accretion flows around Kerr black holes
}

\author[ Pal et al.]{
Kalyanbrata Pal,$^{1,2}$\thanks{E-mail: kalyanbratapal@hri.res.in },
Souvik Ghose$^{1}$\thanks{E-mail: dr.souvikghose@gmail.com},
Shilpa Sarkar,$^{1}$\thanks{E-mail: shilpa.sarkar30@gmail.com}
and
Tapas K. Das$^{1}$\thanks{E-mail: tapas@hri.res.in}
\\
$^{1}$Harish-Chandra Research Institute (HRI),  Chhatnag Road, Jhunsi, Prayagraj (Allahabad), 211019, India. \\
$^{2}$Homi Bhabha National Institute (HBNI), Training School Complex, Anushakti Nagar, Mumbai,Maharashtra 400094, India
}

\date{Accepted XXX. Received YYY; in original form ZZZ}

\pubyear{\the\year{}}

\begin{document}
\label{firstpage}
\pagerange{\pageref{firstpage}--\pageref{lastpage}}
\maketitle

\begin{abstract}
We investigate the axially symmetric accretion of low angular 
momentum hydrodynamic matter onto a rotating black hole. The 
gravitational field under consideration is assumed to be described by a 
pseudo-Newtonian Kerr potential. The accreting matter consists of different species defined by a relativistic equation of state with a variable adiabatic index.We construct 
and solve the hydrodynamical conservation equations governing such a flow, and 
find out the corresponding stationary integral solutions. We find that 
depending on the values of initial boundary conditions, accretion flow 
may exhibit multi-transonic behaviour, and a standing shock may form. We 
investigate, in minute detail, how the spin angular momentum of the 
black hole, as well as the composition of the accreting matter influence the dynamics of accretion flow and the astrophysics of shock formation in the aforementioned accreting 
black hole systems.
\end{abstract}




\section{Introduction}
Accretion is a process by which an astrophysical object accumulates matter from its surroundings, due to its strong gravitational potential energy. 
This process is one of the most important mechanisms which could explain various high-energy astrophysical phenomena of compact objects observed in the Universe. The extreme emissions from gamma-ray bursts (GRB), X-ray binaries (XRB), Active-Galactic Nuclei (AGN) etc. could be explained using this phenomenon. 
Gravitational energy extracted from the in-falling matter powers the aforementioned objects \citep{2002apa..book.....F}. In the last five decades, various models of accretion disks have been proposed, which could explain various aspects of the accretion flow. The basic framework for any disk model, involves, solving hydrodynamical (in the presence of magnetic field magneto-hydrodynamical) equations for the accreting matter, in the background of black hole (BH) spacetime. However, solving these equations is not trivial and hence from time to time, accretion models were progressively developed to explain the present state-of-the-art observations.

One of the first accretion models was given by \citet{1952MNRAS.112..195B} who computed the solutions for spherical accretion onto a compact star. This model provided one of the cornerstones in the theory of accretion and is still now used to explain the emissions from compact objects. However, in this model, the intrinsic angular momentum of the accreting matter was absent. In realistic astrophysical scenarios, it has been found that the accreting matter generally possess some angular momentum, thus, matter would not fall spherically but will spiral inwards, forming an accretion disk-like structure. This theory was proposed by \cite{1973A&A....24..337S} (SS73 hereafter) and the accretion disks thus formed are famously known as the \citeauthor{1973A&A....24..337S} disks.
This disk model also known as the standard disk model of BH accretion, considered a geometrically thin disk and adopted a Newtonian potential. 
Although this disk model was very good at explaining the thermal component part of the emitted radiation, 
\citep{1981ARA&A..19..137P, 2006ARA&A..44...49R, 2008bhad.book.....K} but it was unable to explain the hard non-thermal power-law component of the spectra. Additionally, matter flow in this model was chosen to be Keplerian, cold (temperatures varied around $10^4-10^7K$), optically thick and radiatively efficient. These assumptions need not be true in realistic scenarios. SS73 also neglected the inward radial velocity, with the inner boundary arbitrarily truncated at the inner stable circular orbit (ISCO). 
To explain the non-thermal component, radiatively inefficient, hot advection-dominated accretion flow (ADAF) models were proposed which assumed optically thin gas. The heat generated inside the flow was assumed to be advected inwards with the flow, towards the central object rather than being completely radiated away out of the system, as was in the case of SS73. These ADAF 
were investigated in detail by many authors. Some of the important being by \citet{1976ApJ...204..187S, 1977ApJ...214..840I, 1982Natur.295...17R, 1995ApJ...438L..37A, 1995ApJ...443L..61C, ny95, abra96, esin96}   (see \cite{abra13} for a review).
\citet{1994ApJ...428L..13N, ny95} extended the ADAF models and studied detailed emission processes from accretion flows around compact objects like BHs and neutron stars (NSs). However, they assumed self-similarity which fails to describe the dynamics of flows, especially near the horizon, which is of major importance (\citealt{yn14}).  It was immediately realised that because of the inner boundary condition imposed by the event horizon, BH accretion should be necessarily transonic in nature, 
\citep{1980ApJ...240..271L}. It was seen that for a certain class of accretion flows, especially with low angular momentum, such sonic state transition may take place more than once, and one obtains multi-transonic, axially-symmetric accretion of hydrodynamic fluid onto astrophysical BHs. Such multi-transonic accretion may be endowed with a steady, standing, stationary shock 
\citep{1981ApJ...246..314A, Fukue1983, Lu1985, Lu1986, Fukue1987, 1987MNRAS.227..975B, 1989ApJ...347..365C, cbook90, Nakayama1994, Yang1995, Chakrabarti1996, Pariev1996, Lu1997, Peitz1997, 2001ApJ...557..983D, Barai2004, Takahashi2007, Nagakura2008, Nagakura2009,Das2012,ketal13,kc14,Tarafdar2015,Sukova2015,ck16,Le2016,Sukova2017,kc17,Palit2019,sc19a,sc19b,Palit2020b,scp20,Tarafdar2021,sc22}. Study of shocked flow helps to understand the spectral signature of the BH candidates (\citealt{Chakrabarti_1995,scp20,sc22} and references therein). These shocks are ubiquitous and are a property of transonic flows around compact objects. Not only around BHs, but shocks also have profound implications on spectral properties around neutron stars (NSs)  \citep{kuldeep18,skcp23}. The strong magnetic field lines channel the accreted matter onto the surface of the NS, where it undergoes a shock transition which helps in radiating away the kinetic energy of the accreted matter. This process is in contrast with BHs, which serve as a sink for the accreted matter.

A stationary, multi-transonic, shocked, integral solution is usually obtained for a steady accretion flow. 
Large-scale astrophysical flows around BHs are, however, vulnerable to perturbative events like star-disk interactions or supernovae explosions (\citealt{ceb3f797cb314022b9dbeca50c69a84a} and references therein). 
One thus performs stability analysis of the aforementioned steady flows to ensure that they are stable under such perturbations and thus we can study stationary flow solutions for various accretion-related phenomena 
\citep{ray2003linearized, ray2007evolution, naskar2007acoustic, bhattacharjee2007secular, 2006MNRAS.373..146C, PhysRevD.98.123022}.

In recent years, it has been observed that perturbation of transonic fluid flow under the influence of strong gravity leads to the emergence of a special kind of space-time metric which describes the propagation of perturbation inside the flowing fluid. For linear perturbation, the propagating waves are sound waves. The aforementioned space-time metric describes the propagation of the acoustic perturbation inside the accretion flow and is called the sonic or acoustic metric. A sonic metric is conformally equivalent to a certain representation of Schwarzschild metric and hence possesses acoustic horizons. Such acoustic horizons can be identified with the sonic surfaces produced in the transonic BH accretion. An accreting BH system can thus be looked upon as an interesting physical configuration where the original BH metric and BH-like acoustic metric co-exist 
\citep{Shaikh_2017, PhysRevD.100.043024, PhysRevD.106.044062}. An accreting compact object, thus, can be studied from various perspectives -- from an astrophysical point of view, using the techniques associated with the dynamical systems theory, as well as a natural example of the classical analogue gravity model (\citealt{ghose2024bondiflowvariousperspectives}).

In this work, we are focused on obtaining correct accretion solutions around rotating BHs or Kerr BHs. 
We are motivated to work on these types of BHs because the region near the event horizon is of considerable importance. Most of the exotic processes occur there. The event horizon of Schwarzschild BHs (BHs with spin 0) is at $\rs=2G\mbh/c^2$, where $G$ is the universal gravitational constant, $\mbh$ is the mass of the BH, and $c$ is the speed of light in vacuum. As the spin of the BH is increased, the event horizon is dragged to a region less than the Schwarzschild radius. This encourages us to study accretion flows around BHs with different spin parameters.

Majority of the works done in the literature used a fixed adiabatic index ($\Gamma$) for the equation of state (EoS) of the fluid under consideration. $\Gamma$ is chosen to be 5/3 for non-relativistic flows and 4/3 for relativistic flows. 
But length scales of accretion flow around AGNs or BHXRBs are very large and as a result, the flow may not be fully non-relativistic or relativistic. In general, flows around BHs are trans-relativistic in nature, $\ie$, non-relativistic very far away and relativistic near the horizon (since the BH boundary condition insists that matter should cross the horizon at the speed of light). 
Thus, a fixed $\Gamma$ throughout the whole length scale of the flow is untenable. Temperatures and velocities also change drastically throughout the process. 
To study these types of flows, we need a relativistic equation of state (REoS) which could accurately describe the fluid dynamics. In this work, we have used the \citet{2009ApJ...694..492C} REoS (CREoS, hereafter), which incorporates variable adiabatic index depending on the relativistic nature of the system. Additionally, this REoS also takes care of flows with different species. Electrons being less massive become relativistic at very low temperatures, while protons being heavier need to reach higher temperatures to become relativistic. CREoS helps to deal with these systems accurately.

Although the EoS as introduced by \cite{2006ApJS..166..410R,2009ApJ...694..492C} has been used to study stationary, transonic, shocked accretion solutions, a detailed stability analysis of such flows has not been performed yet. We plan to perform such a task in a series of papers, where we would study an accreting BH system from various perspectives $\ie$, from astrophysical dynamical systems, as well as analogue gravity point of view, where the flow will be described by the CREoS.
The current work deals with the study of BH accretion governed by the aforementioned EoS which is an extension of the work done by \cite{paul2025gravity}. As an initial part of the project, we plan to study shocked accretion flow for axially symmetric accretion maintained in the hydrostatic equilibrium along the vertical direction. The non-self-gravitating accretion takes place under the influence of gravity as described by post-Newtonian pseudo-Kerr BH potentials as introduced by \cite{1996ApJ...461..565A}. In subsequent works (under preparation) we will present the stability analysis of the corresponding stationary solutions and will investigate the emergence of gravity-like phenomena associated with the construction of the corresponding BH-like sonic metric.

In what follows, we first provide a summary of the main features of the REOS {and nature of the potential} used in this work.  We then discuss the governing equations for fluid flow under consideration and the assumptions that we have made for our study. {Next, we discuss the method used to solve the flow dynamics. We then present the effect of variation of  spin parameter and composition parameter on the flow dynamics in the result and analysis section. Finally, we make our concluding remarks.} Throughout our work, the scaling relations of mass, speed, length, time, angular momentum, energy  are given by $\mbh$, $c$, $GM_\mathrm{BH}/c^2$,   $GM_\mathrm{BH}/c^3$,  $GM_\mathrm{BH}/c$ and $M_\mathrm{BH}c^2$  respectively. We will also adapt $G$ = $c$ = $M_\mathrm{BH}$ = 1 for convenience.






\section{Details of the relativistic equation of state (REoS) used  }




\cite{1939isss.book.....C} gave the first relativistically perfect equation of state which was later modified by \cite{MR0088362} and \cite{1968pss..book.....C} to include it in computations. However, all of these EOSs have modified Bessel functions which are cumbersome to implement in numerical calculations as well as in simulations. Thus, as discussed in the introduction we used the CREoS where $\Gamma$ varies automatically to adjust with the thermodynamic state of the flow.  Additionally, it allows us to use different compositions of the flow. 

Here we will consider that our fluid is composed of electrons ($\rm e^-$), positrons ($\rm e^+$) and ions/protons ($\rm p^+$) while maintaining overall charge neutrality. Thus, we have:
\begin{equation}
    n_{\rm e^-}=n_{\rm e^+}+n_{\rm p^+}
\end{equation}
The mass density ($\rho$) of the flow is given by:
\begin{equation}
    \rho=\Sigma_{\mathrm{i}} n_{\mathrm{i}} m_{\mathrm{i}}=n_{\mathrm{e}^{-}} m_{\mathrm{e}^{-}}[2-\xi(1-1 / \eta)]=\rho_{\mathrm{e}^{-}}\tau.
    \label{density.eq}
\end{equation}
Here, $n_\mathrm{i}$s and $m_{\rm i}$s are the number density (in cm$^{-3}$) and mass (in gm) of the $i$-th species. 
Other definitions are: $\rho_{\rm e^-}=n_{\rm e^-} m_{\rm e^-}$, $\tau=[2-\xi(1-1 / \eta)]$, $\eta=m_{\mathrm{e}^{-}} / m_{\mathrm{p}^{+}}$ and $ \xi=n_{\mathrm{p}^{+}} / n_{\mathrm{e}^{-}}$ is the relative proportion of protons compared with electrons, which is also known as the composition parameter. The thermal pressure ($p$) of the flow is given as:
\begin{equation}
    p=\Sigma_{\mathrm{i}} p_{\mathrm{i}}=2n_{\mathrm{e^-}}k_\mathrm{B}T=2 \rho_{\mathrm{e^-}}c^2 \Theta,
    \label{pressure.eq}
\end{equation}
where, $T$ is the temperature of the flow in Kelvin and the non-dimensional temperature is given as $\Theta = k_\mathrm{B}T/(m_\mathrm{e^-}c^2)$. Here, $k_\mathrm{B}$ is the Boltzmann constant.

The energy density of the CREoS for multi-component fluid, simplified using the above definitions of $\rho$ and $p$ is given by \citep{2009ApJ...694..492C}:
\begin{equation}
    \bar{e}= \frac{\rho c^2  f }{\tau}=\frac{\rho  f }{\tau} ~~~[c=1, \mbox{ for the system of units used}],
\end{equation}
where,
$$f(\Theta) =(2-\xi)\left[1+\Theta\left(\frac{9 \Theta+3}{3 \Theta+2}\right)\right]+\xi\left[\frac{1}{\eta}+\Theta\left(\frac{9 \Theta+3 / \eta}{3 \Theta+2 / \eta}\right)\right]$$
Enthalpy ($h$) of the system is given as:
\begin{equation}
    h(\Theta) =\frac{(\bar{e}+p)}{\rho} =\frac{f+2  \Theta}{\tau}.
\end{equation}
The expressions for the polytropic index ($N$) and the adiabatic index ($\Gamma$) for the CREoS are given by as:
\begin{equation}
    N=\frac{1}{2} \frac{ \der f}{\der \Theta} {~~~~\rm and~~~~~}\Gamma = 1 + \frac{1}{N}
    \label{poly_index.eq}
\end{equation}
\noindent The local sound speed ($c_\mathrm{s}$) is defined as:
\begin{equation}
    c_\mathrm{s}^{2} = \frac{2\Theta \Gamma}{\tau}.
\end{equation}

\section{Nature of the BH potentials}

In this present paper, we will use a pseudo potential proposed by Artemova, Bjornsson and Novikov (hereafter ABN) \citep{1996ApJ...461..565A} to mimic the effects of space-time around the rotating (Kerr) BHs. Free fall acceleration is expressed as: 
\begin{equation}
    F_\mathrm{ABN} = - \frac{1}{r^{2 -\beta}(r - \rp)^{\beta}}
    \label{f_ABN.eq}
\end{equation}
 In the above expression:
    \[ r_{+} = 1 + (1 + a^2)^{1/2},  \] 
     \[Z_{1} = 1 + (1 - a^2)^{1/3}[(1 + a)^{1/3} + (1 -a)^{1/3}],\] 
    \[Z_{2} = (3a^2 + Z_{1}^2)^{1/2},\] 
    \[r_\mathrm{in} = 3 + Z_{2} - [(3 - Z_{1})(3 + Z_{1} + 2Z_{2})]^{1/2}, \]
    \[\beta = \frac{r_\mathrm{in}}{r_{+}} - 1,\] 
where, $a$ is the spin parameter or Kerr parameter and $r$ is the radial coordinate, measured along the equatorial plane. $r_{+}$ is defined as the horizon of the rotating BH and $r_\mathrm{in}$ is the innermost stable circular orbit (ISCO).

Integrating the above expression of free-fall acceleration  with proper boundary conditions (potential should vanish at infinity) we can calculate the form of the pseudo-potential, which can be written as:
\begin{equation}
    \Phi_\text{ABN}=\frac{1}{(1-\beta) \rp}\left(1-\frac{\rp}{r}\right)^{1-\beta}-\frac{1}{(1-\beta) \rp}
    \label{ABN_potential.eq}
\end{equation}

A salient feature of the above potential is that, if we put the value of the spin parameter zero, $\ie$ $a = 0$, then the potential will reduce to the usual \citeauthor{1980A&A....88...23P} potential \citep{1980A&A....88...23P}:
\begin{equation}
   \Phi_\text{PW} = -\frac{1}{(r - 2)}
\end{equation}

\section{Details of the disk structure}

We adopt a cylindrical-polar coordinate system ($r$, $\phi$, $z$) to study the inviscid accretion flow around the rotating BH. 
The present work deals with low angular momentum sub-Keplerian flow where viscosity is not essential to allow the matter to
fall in. In these flows, stable circular orbits do not form. Although not a generic phenomenon, but several astrophysical
systems in nature are believed to be characterized by such low angular momentum flow, the most prominent example
is, perhaps, accretion flow onto our own Galactic centre \cite[and references therein]{2006MNRAS.370..219M}.
Apart from that, there exist semi-detached binary systems fed by accretion from OB stellar winds \citep{1975A&A....39..185I, 1984SSRv...38..353L}
where such inviscid accretion flows are found to be relevant. One also finds such a flow pattern 
for semi-detached low-mass non-magnetic binaries \citep{Bisikalo_1998}
and for super-massive BHs in general which are fed by accretion from slowly rotating central stellar
clusters \citep{kbp, ho1998supermassive}.
One of the prominent effects of having a viscous flow is the decreased value of flow angular momentum towards the horizon. But we see in the works by \cite{ck16,kc17,sc22} that even in the presence  of viscosity, $\lambda$ remains constant for a large region of the accretion flow, starting from around $\sim 100\rs$ till the horizon. Thus, we can expect the presence of viscosity to affect the solutions quantitatively rather than qualitatively. We can then safely ignore the presence of viscous effects, which will complicate the objective at hand, and will attend it in an upcoming work.


The axis of rotation of the BH is along the $z$-direction and the mid-plane of the flow is assumed to be the equatorial plane, $z = 0$ plane. We also assume that the flow is symmetric around the $z$-axis, $\ie$, the flow variables are independent of the $\phi$-coordinate and the fluid is in hydrostatic equilibrium  along the vertical direction ($\ie$, along the $z$ axis). In addition to that we consider that our system is in steady state $\ie$, dynamical and thermodynamical variables are independent of time coordinate ($t$). The evolution of accreting matter is governed by conservation equations.

The mass conservation of the flow is governed by the following equation (for a discussion, see  Appendix~\ref{def_Sigma}):
\begin{equation}
    \frac{\partial \Sigma}{\partial t} + \frac{1}{r}\frac{\partial (\Sigma \vel r)}{\partial r} = 0,
    \label{m_con.eq}
\end{equation}
where, $\vel$ is the inward radial velocity of the fluid. 
Here $\Sigma$ denotes the vertically-integrated surface density, which is defined by: $\Sigma = 2\rho H$ (also see in Appendix~\ref{def_Sigma}) and $H$ is the half-height of the disk, measured from the disk mid-plane as given by (see  Appendix \ref{disk_height}):
\begin{equation}
    H  = c_\mathrm{s}\sqrt{\frac{ r}{\Gamma |F_\mathrm{ABN}|}} =\sqrt{\frac{2\Theta r^{3 - \beta}(r - \rp)^{\beta}}{\tau}},
    \label{height.eq}
\end{equation}
where the sign $|f|$ means  the mod value of $f$.

Under the steady-state assumption, the mass conservation equation~(\ref{m_con.eq}) can be integrated to get the mass accretion rate of the system: 
\begin{equation}
    \dot M = 2\pi\Sigma \vel r.
    \label{mass_accretion_rate.eq}
\end{equation}
This is a constant of motion throughout the flow.


Due to the symmetry of the problem and because of the absence of viscosity, we consider only the radial momentum balance condition, which is given by the radial component of the Euler equation. Under steady-state conditions, it is given as:
\begin{equation}
   \vel \frac{\der \vel}{\der r} + \frac{1}{\rho}\frac{\der p}{\der r} - \frac{\lambda^2}{r^3} - F_\mathrm{ABN} = 0  ,
   \label{eu_steady.eq}
\end{equation}
%
%
where,  $\lambda$ is the specific angular momentum (angular momentum per unit mass) of the flow. This is constant throughout the flow in the absence of viscosity. 


Specific energy ($\mathcal{E}$, energy per unit mass) of the flow is obtained by integrating equation (\ref{eu_steady.eq}) and is defined as:
\begin{equation}
    \mathcal{E} = \frac{\vel^2}{2} + h + \frac{\lambda^2}{2r^2} + \Phi_\mathrm{ABN}.
\end{equation}
$\mathcal{E}$ is also known as the Bernoulli parameter, which is constant along the streamlines of the flow. 

The first law of thermodynamics or the conservation of energy is given as:
\begin{equation}
    \frac{\der e}{\der r} - \frac{p}{\rho^2}\frac{\der \rho}{\der r} =0,
    \label{first_law.eq}
\end{equation}
where, the quantity $e$ is defined as: $e = \bar{e}/\rho = f/\tau$.

The form of entropy accretion rate ($\mathcal{\dot M}$) is defined as (see \citealt{kumar2013effect}):
\begin{equation}
    \entmdot= \text{H}\vel r \exp \left(k_3\right) \Theta^{3 / 2}(3 \Theta+2)^{k_1}(3 \Theta+2 / \eta)^{k_2} ,
\end{equation}
where, $k_1=3(2-\xi) / 4$, $k_2=3 \xi / 4$ and $k_3=(f-\tau) /(2 \Theta)$.
Since, there is no dissipation in the system, $\entmdot$ is a constant of motion throughout the flow, apart from $\mdot$ and $\be$.
Using equation (\ref{mass_accretion_rate.eq}) along with the expressions for $N$ (equation \ref{poly_index.eq}) and $H$ (equation \ref{height.eq}) into equation (\ref{first_law.eq}) and after some manipulation we could write the rate of change in non-dimensional temperature ($\Theta$) with respect to the radial distance as:
\begin{equation}
    \frac{\der \Theta}{\der r} = \Omega_{1} + \Omega_{2}\frac{\der \vel}{\der r},
    \label{dtheta_dr.eq}
\end{equation}
where,
\[\Omega_{1} = -\frac{\Theta}{(2N + 1)} \Big[ \frac{(5 - \beta)}{r} + \frac{\beta}{(r -\rp)} \Big] \],
and,
\[\Omega_{2} = - \frac{2\Theta}{(2N + 1)\vel}        \].


We could rewrite equation (\ref{eu_steady.eq}), using equation (\ref{mass_accretion_rate.eq}) and equation (\ref{dtheta_dr.eq}) in the following way: 
 \begin{equation}
     \frac{\der  \vel}{\der r} = \frac{\frac{c_\mathrm{s}^{2}}{(\Gamma + 1)}\bigg[  \frac{(5 - \beta)}{r} + \frac{\beta}{(r -\rp)}\bigg] + \frac{\lambda^{2}}{r^3} - \frac{1}{r^{2 -\beta}(r - \rp)^{\beta}}}{\bigg[\vel - \frac{\mathrm{c_\mathrm{s}^2}}{\vel}\frac{2}{(\Gamma + 1)}\bigg]} = \frac{\mathcal{N}}{\mathcal{D}}.
     \label{dv_dr}
 \end{equation}
 We need to solve the above two coupled non-linear differential equations (equations~\ref{dtheta_dr.eq} and~\ref{dv_dr}) numerically to construct the radial velocity profile for the inviscid non-dissipative axisymmetric fluid flow around a rotating BH. We also need to specify a set of parameters: [$\mathcal{E}$, $\lambda$, $a$, $\xi$] to solve the dynamical equations. In the next section, we discuss the methodology used to solve the coupled differential equations along with the finding of shock domain in these types of flows. 


\section{Stationary integral solutions of flow dynamics}
In this section, we will start with the conditions for finding the critical point (points) in the flow  and then discuss whether the flow will admit shock or not and its implication for our study.
\subsection{Critical point analysis method}
At the outer boundary, the accreting matter starts with some low radial velocity ($\vel$), which is usually subsonic ($\vel<c_\mathrm{s}$) in nature. As the matter moves inward due to gravitational pull its velocity increases. At the same time, as a result of increasing temperature (due to fluid compression), the local sound speed ($c_\mathrm{s}$) also increases. If at a certain point (points) the fluid velocity crosses the local sound speed and becomes supersonic ($\vel>c_\mathrm{s}$) we call this point (points) a sonic point (points). Due to the inner boundary condition (any matter should cross the BH horizon with the speed of light), the fluid must attain at least one sonic point and the flow is called a transonic flow \citep{1980ApJ...240..271L} and if there are multiple sonic points (MSPs), then the flow is known to be a multi-transonic flow. Before locating the sonic point, we have to calculate the critical point for the flow equation. The location of sonic point (points) and critical point (points) may coincide with each other or they may be different depending on the structure of the flow. 

As the matter flow is assumed to be smooth physically, so if at any point (radial coordinate) the denominator ($\mathcal{D}$) of equation (\ref{dv_dr}) vanishes, the numerator ($\mathcal{N}$) should also vanish there, to keep velocity gradient finite. We call this point the critical point ($\rc$, where `c' stands for critical).


The critical point conditions are described by the following equations:
\begin{equation}
\numc =\denc = 0.
\label{N}
\end{equation}
Using these conditions in equation (\ref{dv_dr}), we get the following two equations, which are satisfied at the critical point, $\rc$:
\begin{equation}
c_\mathrm{sc}^2 = \frac{(\Gamma_\mathrm{c} + 1) \bigg[ \frac{1}{r^{2 -\beta}_\mathrm{c}(r_\mathrm{c} - \rp)^{\beta}} - \frac{\lambda^2}{r_\mathrm{c}^3} \bigg]}{ \bigg[ \frac{(5 - \beta)}{r_\mathrm{c}} + \frac{\beta}{(r_\mathrm{c} -\rp)}\bigg]}.
\end{equation}
and
\begin{equation}
\vel_\mathrm{c} = \sqrt{\frac{2c_\mathrm{sc}^2}{\Gamma_\mathrm{c} + 1}}.
\end{equation}
where the quantities with the subscript `c' denoted their values at the critical point.

Since, the value of $\der \vel/\der r$ at the critical point has a 0/0 form, we have to use L'Hopital rule to compute $\frac{\der \vel}{\der r}\Big\lvert_{\mathrm{c}}$. This is given by:
\begin{equation}
    \frac{\der \vel}{\der r}\Bigg\lvert_{\mathrm{c}} = \frac{\frac{\der \mathcal{N}}{\der r}\Big\lvert_{\mathrm{c}}}{\frac{\der \mathcal{D}}{\der r}\Big\lvert_{\mathrm{c}}}
    \label{lhopital_derivative}
\end{equation}
The generic form of $\frac{\der\mathcal{N}}{\der r}$ could be written as:
\begin{equation}
    \frac{\der\mathcal{N}}{\der r} = \mathcal{N}_{1} + \mathcal{N}_{2}\frac{\der \vel}{\der r}
    \label{dN_dr}
\end{equation}
where
\[ \mathcal{N}_{1} = \mathcal{N}_{11} + \Omega_{1}\mathcal{N}_{12}  \] 
\[ \mathcal{N}_{2} = \Omega_{2}\mathcal{N}_{12} \]
Similarly the generic form of $\frac{\der \mathcal{D}}{\der r}$ could be written as:
\begin{equation}
    \frac{\der \mathcal{D}}{\der r} = \mathcal{D}_{1} + \mathcal{D}_{2}\frac{\der \vel}{\der r}
    \label{dD_dr}
\end{equation}
where
\[ \mathcal{D}_{1} = \Omega_1\mathcal{D}_{12} \]
\[ \mathcal{D}_{2} =  \mathcal{D}_{11} + \Omega_{2}\mathcal{D}_{12} \]
In the above expressions, the functional forms of $\mathcal{N}_{11}$, $\mathcal{N}_{12}$, $\mathcal{D}_{11}$ and $\mathcal{D}_{12}$ are presented in  Appendix \ref{coefficients}.

Using equation (\ref{dN_dr}) and (\ref{dD_dr}) in equation (\ref{lhopital_derivative}) and after doing some manipulation the equation (\ref{lhopital_derivative})  could be structured in the following form :
\begin{equation}
    \mathcal{A}\bigg(\frac{\der \vel}{\der r}\bigg)^2\Bigg\lvert _{\rm c}+ \mathcal{B}\bigg(\frac{\der \vel}{\der r}\bigg)\Bigg\lvert _{\rm c} + \mathcal{C} = 0
    \label{root_equation}
\end{equation}
The expression for $\mathcal{A}$, $\mathcal{B}$, $\mathcal{C}$ are given below:
\[\mathcal{A} = \mathcal{D}_{11} +\Omega_{2}\mathcal{D}_{12} \]
\[ \mathcal{B} = \Omega_{1}\mathcal{D}_{12} -\Omega_{2}\mathcal{N}_{12}  \]
\[ \mathcal{C} = -\mathcal{N}_{11} - \Omega_{1}\mathcal{N}_{12} \]

Equation (\ref{root_equation}) is a quadratic equation, which has two roots and given by:
\begin{equation}
   \frac{\der \vel}{\der r}\Bigg\lvert_{\mathrm{c}}= \frac{-\mathcal{B} \overset{+}{-} \sqrt{\mathcal{B}^2 -4\mathcal{A}\mathcal{C}}}{2\mathcal{A}}
    \label{slope}
\end{equation}
The negative root in the equation (\ref{slope}) corresponds to a accretion-type flow while the positive root corresponds to wind solutions.
For global accretion solution, depending on the initial condition, there may be either one critical point or three critical points.
For three critical points, critical point close to the BH horizon is known as the inner critical point ($r_\mathrm{in}$), that formed far away from the horizon is called the outer critical point ($r_\mathrm{out}$) and the one formed in between the inner and outer critical point, is called the middle critical point ($r_\mathrm{mid}$). 
It is to be noted that stationary integral global flows can only pass through a saddle-type critical point. Thus, matter can passes through either $r_\mathrm{in}$ and $r_\mathrm{out}$.
On the other hand, $r_\mathrm{mid}$ is a center-type critical point through which real flow cannot pass. 
For flows containing MCP, the global solution can pass through both the sonic points via a shock transition. Shocks happen in certain sections of the parameter space and has serious implication in astrophysical systems.

\subsection{Shock analysis method}
As discussed before, in MCP regime, shocks may form. In such a case, the flow starting subsonically from the outer boundary becomes supersonic after crossing the $r_\mathrm{out}$. If the shock conditions are satisfied, they undergo a shock transition after which they again become subsonic. However, to satisfy the inner boundary conditions, they pass through the inner sonic point and enter the BH supersonically. 
The state change from supersonic to subsonic through a shock transition could be mediated by the centrifugal barrier due to the rotating motion of the fluid. Through shocks the fluid can change its dynamical or thermodynamical properties discontinuously. Depending on the strength of the barrier the shock could halt the flow of the matter to pass it from its supersonic to subsonic state. At the shock transition the flow should obey the following conditions, known as Rankine-Hugoniot (RH) shock conditions:
\begin{equation}
    \dot M_{+} = \dot M_{-} ,
    \label{m_shock.eq}
\end{equation}
\begin{equation}
    W_{+} + \Sigma_{+}\vel_{+}^2 = W_{-} + \Sigma_{-}\vel_{-}^2 ,
    \label{mom_shock.eq}
\end{equation}
\begin{equation}
    \mathcal{E}_{+} = \mathcal{E}_{-},
    \label{energy_shock.eq}
\end{equation}
where $W = 2Hp$ is the vertically integrated thermal pressure and 
subscripts $(-)$ denotes pre-shocked quantities and $(+)$ denotes post-shocked quantities. Manipulating equation (\ref{m_shock.eq}) we get:
\begin{equation}
    \rho_{+}\vel_{+}H_{+} = \rho_{-}\vel_{-}H_{-}
\end{equation}
and from equation (\ref{mom_shock.eq}) we find:
\begin{equation}
  H_{+}p_{+} + H_{+}\rho_{+}\vel_{+}^2 = H_{-}p_{-} + H_{-}\rho_{-}\vel_{-}^2  ,
\end{equation}
%

Shock conditions give us the above mentioned quantities which are conserved throughout the supersonic and subsonic branch. To compute the shock location ($r_\mathrm{sh}$), we have to define a quantity, the shock invariant quantity ($I$), which is the same only at the shock location for both the super and subsonic branches of the flow respectively. Using the shock condition equations and after some manipulation, we could define the invariant quantity as follows (see Appendix \ref{shock_invariant}):
\begin{equation}
    I = \frac{[\Theta + \frac{\tau}{2}\vel^2]}{\vel}.
\end{equation}
Therefore using the below equation:
\begin{equation}
     \frac{[\Theta_{+} + \frac{\tau}{2}\vel^2_{+}]}{\vel_{+}} = \frac{[\Theta_{-} + \frac{\tau}{2}\vel^2_{-}]}{\vel_{-}}.
\end{equation}
we can calculate the shock location.


\begin{figure*}
    \centering
    \includegraphics[width=1\linewidth]{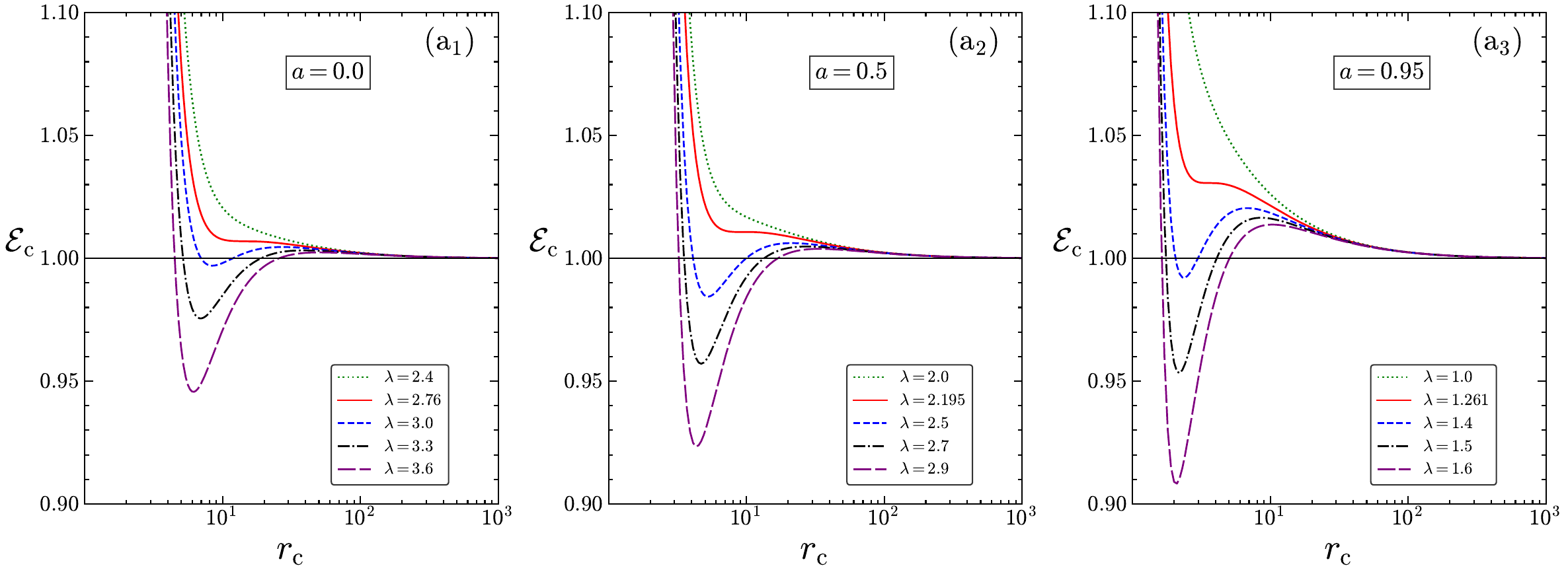}
    \caption{The critical specific energy ($\mathcal{E}_\text{c}$) is plotted against the location of critical points ($r_\text{c}$) for (a$_1$) $a$ = 0.0, (a$_2$) $a$ = 0.5 and (a$_3$) $a$ = 0.95, considering various values of specific angular momentum ($\lambda$). Specific angular momentum corresponding to each curve is as follows: (a$_1$) $\lambda$ = 2.4 (dotted), $\lambda$ = 2.76 (solid), $\lambda$ = 3.0 (dashed), $\lambda$ = 3.3 (dashed dotted) and $\lambda$ = 3.6 (long dashed), (a$_2$) $\lambda$ = 2.0 (dotted), $\lambda$ = 2.195 (solid), $\lambda$ = 2.5 (dashed), $\lambda$ = 2.7 (dashed dotted) and $\lambda$ = 2.9 (long dashed), (a$_3$) $\lambda$ = 1.0 (dotted), $\lambda$ = 1.261 (solid), $\lambda$ = 1.4 (dashed), $\lambda$ = 1.5 (dashed dotted) and $\lambda$ = 1.6 (long dashed).
    The number of intersection points between the horizontal line plotted at $\mathcal{E}_\mathrm{c} = 1.0001$ in each diagram and the curves represents the number of critical points for that particular flow. Curves corresponding to limiting value of specific angular momentum ($\lambda$) are shown as solid lines. In all cases, the composition parameter is set to $\xi = 1.0$.  }
    \label{fig:1}
\end{figure*}

\section{Results and Analysis}

In this section, we discuss the key results of our study. First, in Subsection 6.1, we study the location and number of critical points for various values of specific energy ($\mathcal{E}$) and specific angular momentum ($\lambda$). We also construct the energy momentum parameter space ($\mathcal{E}-\lambda$), which allows us to predict whether the flow will exhibit single or multiple critical points based on the flow parameters. For demonstration, we consider three typical values of the BH spin parameter: $a = 0.0$, $0.5$ and $0.95$, where $a = 0.0$ corresponds to a non-rotating BH.
In Subsection 6.2, we investigate the formation of shocks, the associated shock parameter space, and how global multi-transonic accretion solutions arise from multi-critical accretion flows. Next, in Subsection 6.3, we examine the effect of the BH spin parameter ($a$) on the shock location ($r_{\text{sh}}$) and other shock-induced flow variables. Finally, in Subsection 6.4, we explore how variations in the composition parameter ($\xi$) influence the multi-critical parameter space (a subregion in the ($\mathcal{E}-\lambda$) space that contains multiple critical points) and the shock parameter space.
In addition to that, we investigate the dependence of the shock location ($r_{\text{sh}}$) and related quantities on the composition parameter ($\xi$) of the accretion flow.


\subsection{Critical point analysis and multi-critical parameter space}

In this subsection, we analyze how the critical specific energy ($\mathcal{E}_{\text{c}}$), $\ie$, specific energy at the critical point of the flow varies with the critical points ($r_{\text{c}}$) for different values of the BH spin parameter ($a$) and specific angular momentum ($\lambda$) of the flow. We consider the composition parameter $\xi = 1.0$, corresponding to a pure electron-proton ($e^- - p^+$) flow.  

In Fig.~(\ref{fig:1}a$_1$), we plot the critical specific energy ($\mathcal{E}_{\text{c}}$) as a function of the critical points ($r_{\text{c}}$) for the spin parameter $a = 0.0$. The chosen values of the specific angular momentum are: $\lambda = 2.4$ (dotted line), 2.76 (solid line), 3.0 (dashed line), 3.3 (dashed-dotted line), and 3.6 (long-dashed line). There exists a limiting value of the specific angular momentum ($\lambda = 2.76$) below which the flow possesses only a single critical point for all accessible values of the critical specific energy ($\mathcal{E}_{\text{c}}$).  
In Fig.~(\ref{fig:1}a$_2$), $\mathcal{E}_{\text{c}}$ is plotted against $r_{\text{c}}$ for the spin parameter $a = 0.5$, with corresponding values of the specific angular momentum: $\lambda = 2.0$ (dotted line), 2.195 (solid line), 2.5 (dashed line), 2.7 (dashed-dotted line), and 2.9 (long-dashed line). Here, the limiting value of the specific angular momentum is $\lambda = 2.195$.  
Fig.~(\ref{fig:1}a$_3$) illustrates the variation of $\mathcal{E}_{\text{c}}$ with $r_{\text{c}}$ for the spin parameter $a = 0.95$, where the specific angular momentum values are chosen as $\lambda = 1.0$ (dotted line), 1.261 (solid line), 1.4 (dashed line), 1.5 (dashed-dotted line), and 1.6 (long-dashed line). The limiting value of the specific angular momentum in this case is $\lambda = 1.261$.  
From the figures (Fig.~\ref{fig:1}), it can be observed that the limiting value of the specific angular momentum, below which the flow exhibits only a single critical point, decreases with increasing BH spin parameter. Additionally, as the specific angular momentum ($\lambda$) increases for a fixed spin parameter, the number of turning points increases from one to three. For instance, if a horizontal line is drawn at a fixed critical specific energy value ($\mathcal{E}_{\text{c}} > 1.0$), it will intersect the critical energy curve at one to three points. For demonstration, we have drawn a horizontal line at $\mathcal{E}_{\text{c}} = 1.0001$. The locations of these intersection points correspond to the critical points of the flow for the given values of specific angular momentum and BH spin parameter.

\begin{figure*}
    \centering
    \includegraphics[width=1\linewidth]{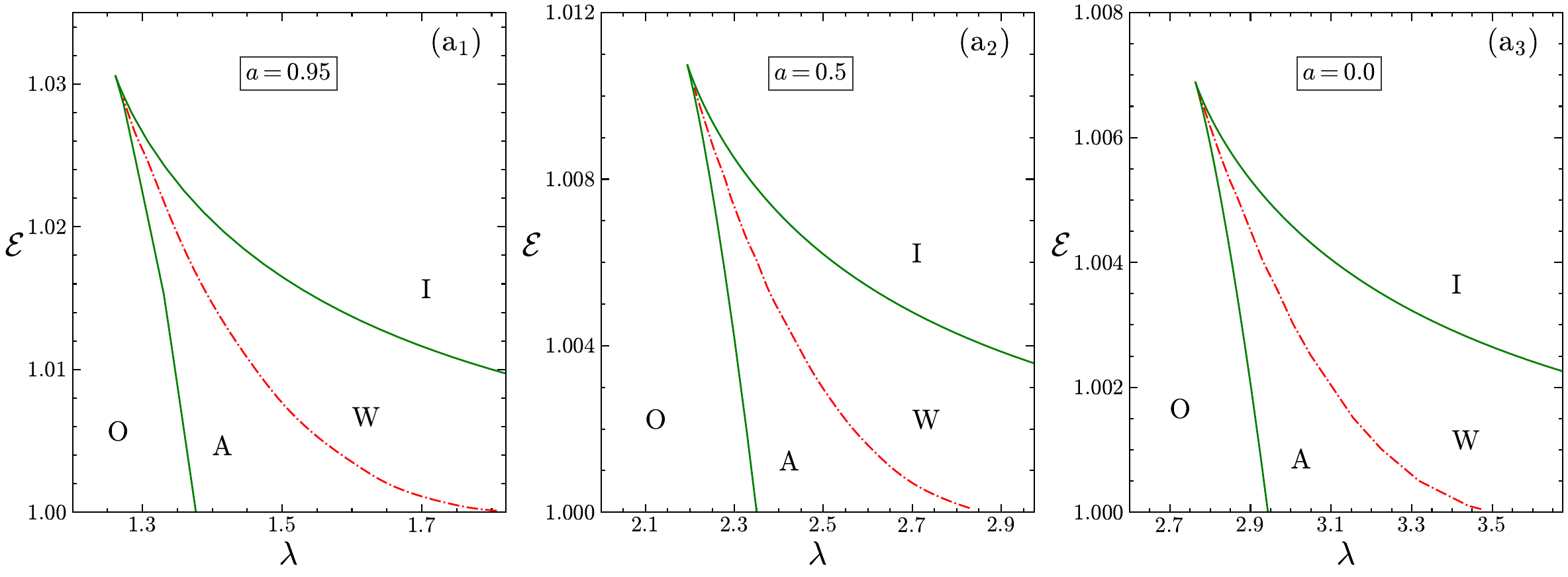}
    \caption{ The energy-momentum parameter space ($\mathcal{E} - \lambda$) is plotted for BH spin parameter value: (a$_1$) $a$ = 0.95, (a$_2$) $a$ = 0.5 and (a$_3$) $a$ = 0.0. Each diagram is divided into four distinct regions: $\boldsymbol{\textbf{O}}$, $\boldsymbol{\textbf{A}}$, $\boldsymbol{\textbf{W}}$, and $\boldsymbol{\textbf{I}}$, corresponding to different phase portrait topologies (see text for details). Region bounded by the solid green line, also known as multi-critical parameter space, contains three critical points, while region outside this contains single critical point. $\xi = 1.0$ for all the above diagrams.  }
    \label{fig:2}
\end{figure*}

Fig.~\ref{fig:1} provides an intuitive understanding of the range of the flow parameters [$\mathcal{E}$, $\lambda$] that determine whether the accretion flow exhibits single or multiple critical points. However, to unveil the complete picture of the problem, it is essential to explicitly construct the energy momentum parameter space ($\mathcal{E} - \lambda$).

In Fig.~\ref{fig:2}, we present the ($\mathcal{E} - \lambda$) parameter space for three BH spin values: $a = 0.95$ (Fig.~2a$_1$), $0.5$ (Fig.~2a$_2$) and $0.0$ (Fig.~2a$_3$). Each parameter space is categorized  into four distinct regions, labeled as $\boldsymbol{\textbf{O}}$, $\boldsymbol{\textbf{A}}$, $\boldsymbol{\textbf{W}}$, and $\boldsymbol{\textbf{I}}$, based on the flow topology and the number of critical points.  
Parameters [$\mathcal{E}$, $\lambda$] chosen from the $\boldsymbol{\textbf{O}}$ or $\boldsymbol{\textbf{I}}$ regions correspond to flows with a single saddle-type critical point. Specifically, the $\boldsymbol{\textbf{O}}$ region represents flows with an \textbf{O}uter critical point ($r_\text{out}$), located far away from the BH horizon, whereas the $\boldsymbol{\textbf{I}}$ region corresponds to flows with an \textbf{I}nner critical point ($r_\text{in}$), situated close to the horizon.  
In contrast, parameter values within the $\boldsymbol{\textbf{A}}$ or $\boldsymbol{\textbf{W}}$ regions produce flows with three critical points: two saddle-type critical points and one center-type critical point. In addition to that, saddle-type critical points are located near the horizon (inner critical point, $r_\text{in}$) and far from the horizon (outer critical point, $r_\text{out}$), while the center-type critical point, situated between the two saddles, is known as the middle critical point ($r_\text{mid}$). Global solution that connects infinity to horizon can not pass through the centre type critical point, so in the context of matter flow $r_\text{mid}$ does not have much relevance.
Both the $\boldsymbol{\textbf{A}}$ and $\boldsymbol{\textbf{W}}$ regions of the parameter space contain three critical points each; however, they can be distinguished by comparing the entropy at the inner and outer critical points. In region $\boldsymbol{\textbf{A}}$, the entropy satisfies:~$\mathcal{\dot{M}}(r_\text{in}) > \mathcal{\dot{M}}(r_\text{out})$, while in region $\boldsymbol{\textbf{W}}$:~ $\mathcal{\dot{M}}(r_\text{in}) < \mathcal{\dot{M}}(r_\text{out})$. {Therefore, for the solution in $\boldsymbol{\textbf{A}}$ region, only the $\boldsymbol{\textbf{A}}$ccretion type flow which passes through the outer sonic point may experience the shock and becomes a multi-transonic flow.} For the solutions in $\boldsymbol{\textbf{W}}$ region only the $\boldsymbol{\textbf{W}}$ind type flow may experience the shock and become a multi-transonic flow. This depends on the satisfaction of the Rankine-Hugonoit conditions. The red dashed-dotted line in Fig.~\ref{fig:2}, which marks the boundary between the two regions, represents the condition where the entropy at the inner and outer critical points is equal.

\begin{figure}
    \centering
    \includegraphics[width=1\linewidth]{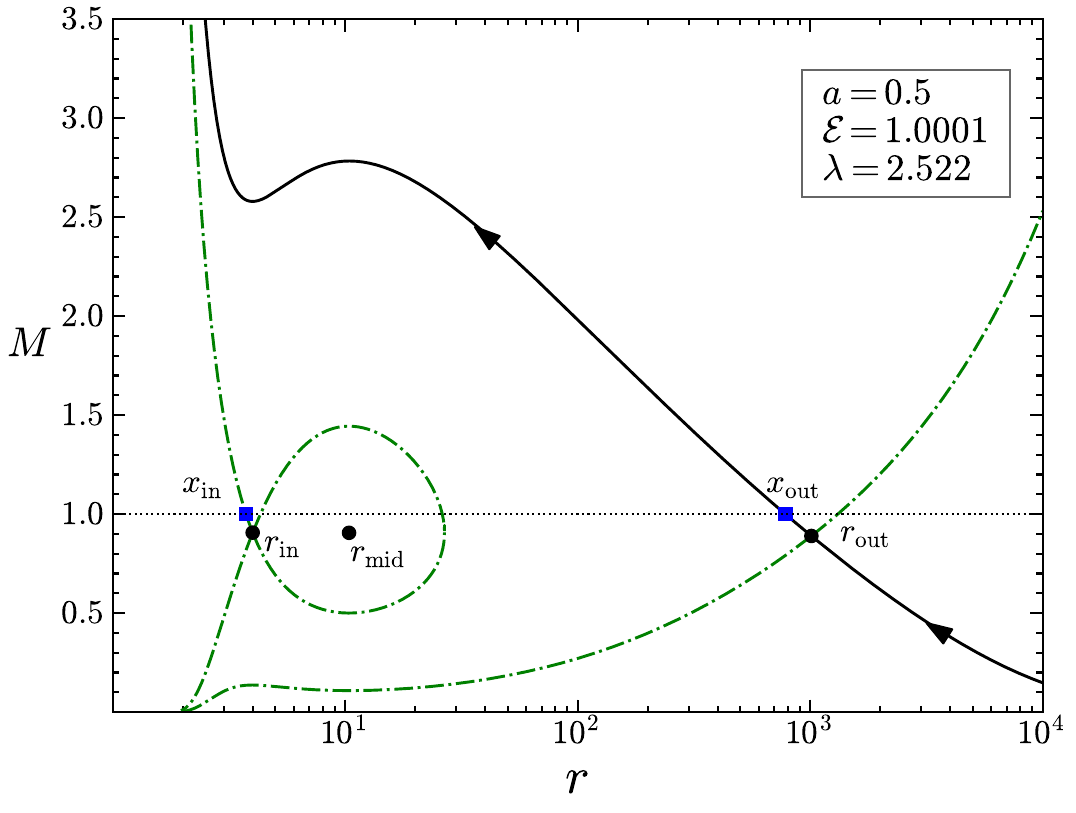}
    \caption{Phase portrait is shown for parameter values: [$\mathcal{E}$, $\lambda$, $a$, $\xi$] = [1.0001, 2.522, 0.5, 1.0]. The solid black line with arrows represents the global accretion solution. Critical points are marked with solid black circles, while sonic points are indicated by solid blue squares. The flow passes only through the outer sonic point before reaching the horizon, making it a mono-transonic flow. }
    \label{fig:3}
\end{figure}

Now we explore the accretion solution in terms of drawing the phase portrait, a Mach number ($ M = \frac{\vel}{c_\mathrm{s}}$) vs radial distance ($ r $) plot, by solving the pair of dynamical equations~(\ref{dtheta_dr.eq}) and~(\ref{dv_dr}) for BH spin parameter value, $a$ = 0.5 and composition parameter, $\xi$ = 1.0. Fig.~(\ref{fig:3}) shows the corresponding phase diagram with $ [\mathcal{E}, \lambda]_{\boldsymbol{\textbf{A}}} $ = [1.0001, 2.522], where the subscript $\boldsymbol{\textbf{A}}$ means the chosen parameter values are from the $\boldsymbol{\textbf{A}}$ region of the $(\mathcal{E}-\lambda)$ parameter space (see Fig.~2a$_2$). In general, for parameter values chosen from the $\boldsymbol{\textbf{R}}$ region we use the $[\mathcal{E}, \lambda]_{\boldsymbol{\textbf{R}}}$ notation. It should be noted that: $[\mathcal{E}, \lambda] \subset [\mathcal{E}, \lambda, a, \xi] $ and here we have already fixed the value of $a$ and $\xi$. For the aforementioned parameter values, the flow dynamics show three critical points which is obvious. Additionally, it is very interesting to note that sonic points location (inner sonic point ($x_{\text{in}}$) and outer sonic point ($x_{\text{out}}$), marked with solid blue square) are not same with that of the critical points location ($r_\text{in}$, $r_\text{out}$). In Fig.~(\ref{fig:3}), the solid black line shows the flow line of matter being accreted by the BH. This line (solid black with arrow) connects the horizon to infinity and it passes only through the outer critical point ($r_{\text{out}}$) as well as through the outer sonic point ($x_{\text{out}}$). Therefore, two parts of the phase diagram, one part which goes through the outer sonic point (solid black) and another part that contains the inner sonic point (green dashed-dotted part) are disconnected. So though the flow geometry shows multi-criticality behaviour,  it is a mono-transonic (passes through a single sonic point, in this case, which is the outer sonic point) flow.  For multi-transonic flow, accreting matter which is essentially subsonic at infinity,  becomes supersonic after crossing the outer sonic point and should go through a state transition from supersonic to subsonic again to accommodate the inner sonic point also and the state change has interceded through a shock. In the next subsection, we inspect how the shock forms in rotating axisymmetric flow around a rotating BH and its implication on multi-critical flow to make it a multi-transonic flow.

\begin{figure*}
    \centering
    \includegraphics[width=1\linewidth]{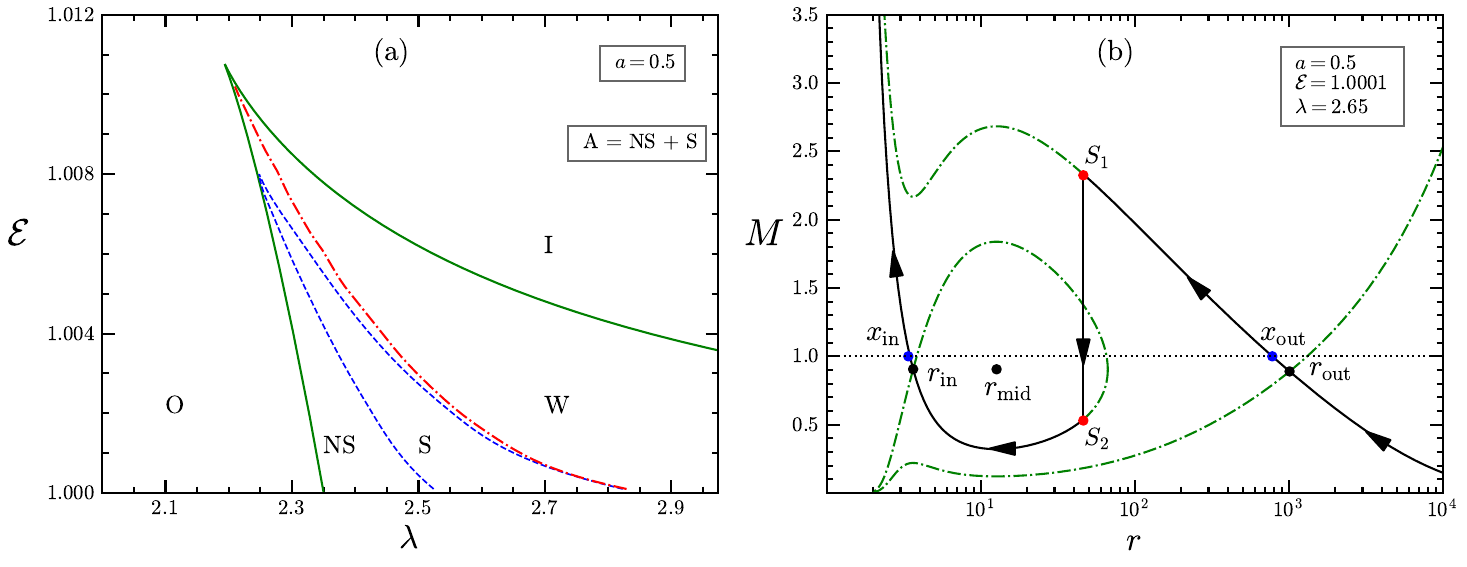}
    \caption{In the left panel we have plotted the shock parameter space (region bounded by the dashed blue line and denoted as $\textbf{S}$) along with the multi-critical parameter space (region inside the solid green line) for $a = 0.5$ and $\xi = 1.0$. In the right panel phase portrait is drawn for parameter values: [$\mathcal{E}$, $\lambda$, $a$, $\xi$] = [1.0001, 2.65, 0.5, 1.0]. Solid black line with arrow represents the global multi-transonic accretion flow solution. Line connected between the points $S_1$ and $S_2$, shows the shock transition in the flow.  }
    \label{fig:4}
\end{figure*}

\begin{figure}
    
    \centering
    \includegraphics[width=1\linewidth]{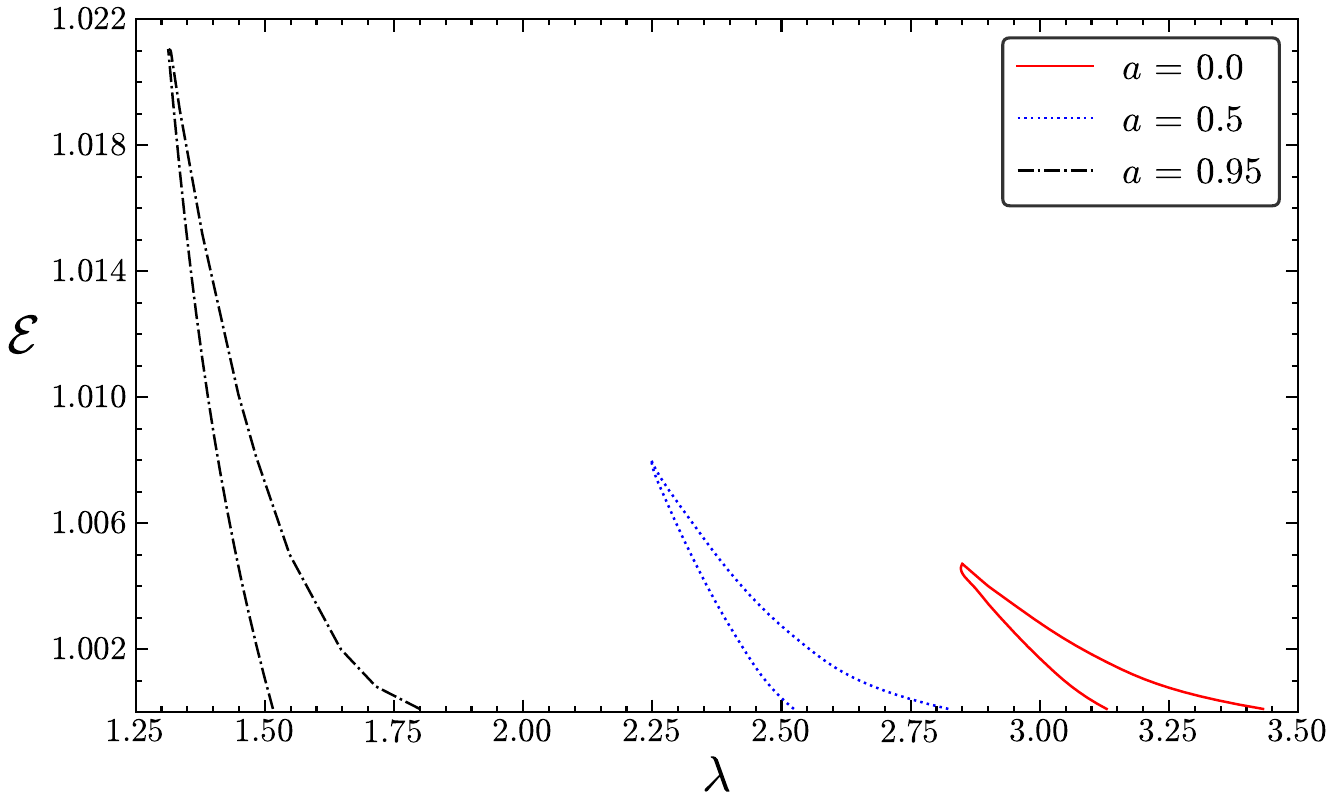}
    \caption{ The region bounded by the curves represents the part of the ($\mathcal{E} -\lambda$) parameter space for which the flow contains shock, also known as shock space. The region bounded by black dashed-dotted line represent shock space for $a = 0.95$, while region bounded by blue dotted line is the shock space for $a = 0.5$ and the region enclosed by solid red line is the shock space for $a = 0.0$.  }
    \label{fig:5}
\end{figure}

\begin{figure*}
    \centering
    \includegraphics[width=0.8\linewidth]{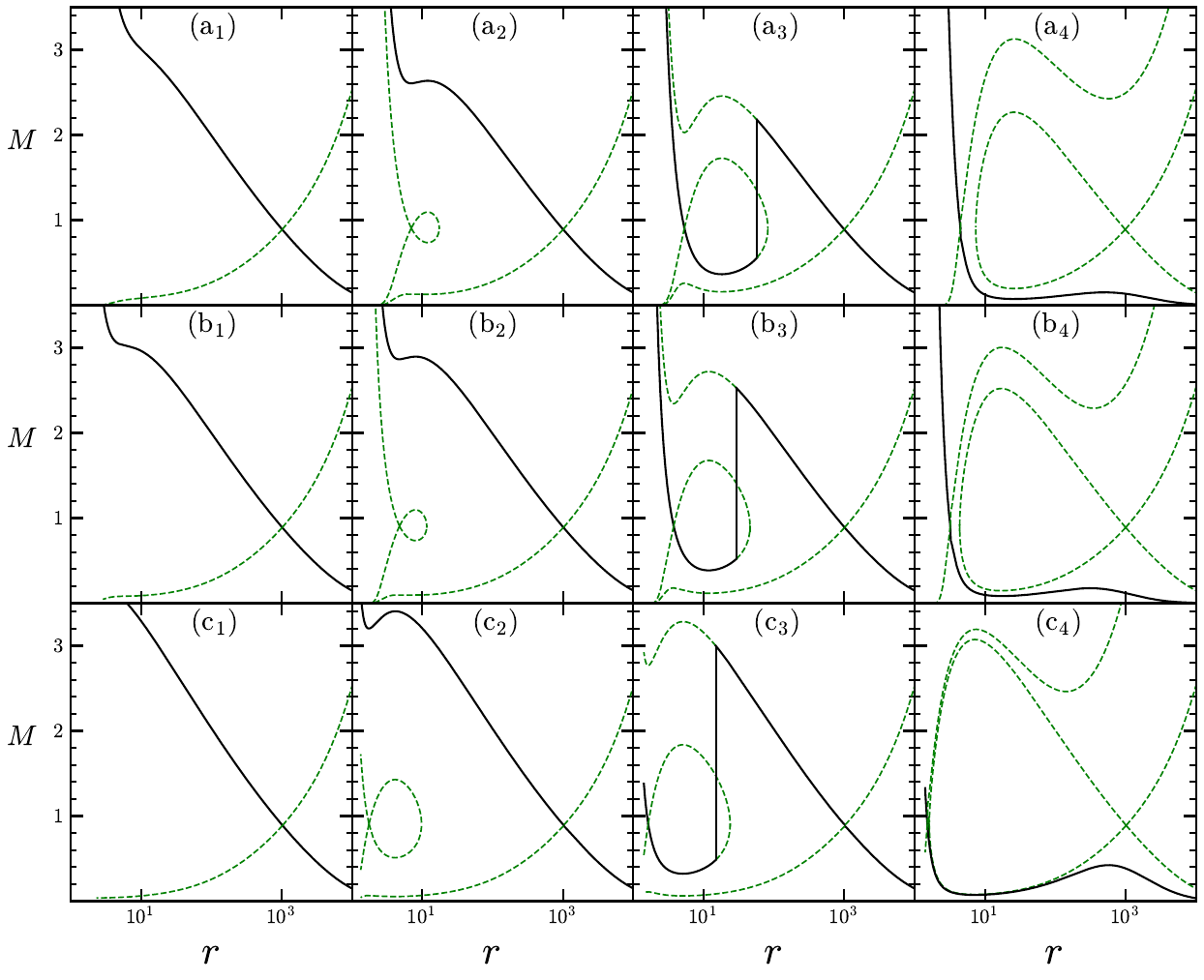}
    \caption{In the upper panel various types of phase portrait ($M$ vs $r$) are being drawn with BH spin parameter $a = 0.0$ for (a$_1$) [$\mathcal{E}$, $\lambda$] = [1.0001, 2.6], (a$_{2}$) [$\mathcal{E}$, $\lambda$] = [1.0001, 3.0], (a$_{3})$ [$\mathcal{E}$, $\lambda$] = [1.0001, 3.25] and (a$_{4})$ [$\mathcal{E}$, $\lambda$] = [1.0001, 3.6].
    In the middle panel phase portrait are being drawn with spin $a = 0.5$ for (b$_{1})$ [$\mathcal{E}$, $\lambda$] = [1.0001, 2.3], (b$_{2})$ [$\mathcal{E}$, $\lambda$] = [1.0001, 2.4], (b$_{3})$ [$\mathcal{E}$, $\lambda$] = [1.0001, 2.6] and (b$_{4})$ [$\mathcal{E}$, $\lambda$] = [1.0001, 2.9]. 
    In the last panel phase portrait are being drawn with spin $a = 0.95$ for (c$_{1})$ [$\mathcal{E}$, $\lambda$] = [1.0001, 1.3], (c$_{2})$ [$\mathcal{E}$, $\lambda$] = [1.0001, 1.5], (c$_{3})$ [$\mathcal{E}$, $\lambda$] = [1.0001, 1.6] and (c$_{4})$ [$\mathcal{E}$, $\lambda$] = [1.0001, 1.81]. For all the above figures we have taken $\xi = 1.0$.
    }
    \label{fig:6}
\end{figure*}


\subsection{Shock formation and multi-transonic accretion solution}

As we have discussed in subsection {(5.2)}, once the accreting fluid becomes supersonic (after crossing the outer sonic point) it may face shock due to the outward centrifugal force and change its state from supersonic to subsonic discontinuously, also known as a shock transition. As a result of this shock transition, the fluid will pass through the inner sonic point also, before reaching the BH horizon, which makes it a multi-transonic accretion flow. Here we have investigated the shock-induced phase portrait and shock parameter space, a subdomain of the ($\mathcal{E}-\lambda$) parameter space which harbours the shock, in detail.

In Fig.~\ref{fig:4}a, we have plotted the shock parameter space, where the flow undergoes a shock transition, along with the ($\mathcal{E}-\lambda$) parameter space for the spin parameter, $a= 0.5$ and composition parameter, $\xi = 1.0$. We have divided the $\boldsymbol{\textbf{A}}$ region (see Fig.~\ref{fig:2}a$_2$) into two subregions: $\boldsymbol{\textbf{NS}}$ (\textbf{N}o \textbf{S}hock region) and $\boldsymbol{\textbf{S}}$ (\textbf{S}hock region), while the other regions remain the same as in Fig.~\ref{fig:2}a$_2$. Consequently, the flow will be monotransonic if the chosen parameter values lie in the $\boldsymbol{\textbf{NS}}$ region: $[\mathcal{E}, \lambda] \in [\mathcal{E}, \lambda]_{\textbf{NS}}$, and it will be multi-transonic for parameters value picked up from the $\boldsymbol{\textbf{S}}$ region: $[\mathcal{E}, \lambda] \in [\mathcal{E}, \lambda]_{\textbf{S}}$.

Fig.~\ref{fig:4}b, shows the global solution for an axially-symmetric multi-transonic shocked accretion flow around a rotating BH. The phase diagram is being constructed for the spin parameter $a = 0.5$ and the composition parameter $\xi = 1.0$ with other parameters chosen from the shock parameter space: $[\mathcal{E}, \lambda]_{\textbf{S}}$ = [1.0001, 2.65]. The solid black line with an arrow, which passes through both the outer and inner sonic points is the multi-transonic accretion flow line. The vertical line $\overline{S_1S_2}$, is the discontinuous flow line by which the sonic points are connected, and represents a shock transition. Due to shock, the flow halts suddenly and as a consequence, its dynamical as well as thermodynamical properties change discontinuously. 

After exploring the significance of shock in multi-transonic accretion flow around a spinning BH, we extend our study to examine how the shock parameter space, a bounded region in the ($\mathcal{E} - \lambda$) parameter space that allows shock formation in accretion flow, varies with changes in the BH spin parameter ($a$). In Fig.~\ref{fig:5}, we have depicted the shock parameter space in the ($\mathcal{E} - \lambda$) plane for spin parameter values, $a$ = 0.0 (region bounded by the solid red line), 0.5 (region bounded by the dotted blue line) and 0.95 (region bounded by the dashed-dotted black line). From the diagram, we notice that: as we increase the spin parameter value ($a$) of the BH, the shock parameter space gets shifted to the lower specific angular momentum side in the ($\mathcal{E} - \lambda$) plane  and specific energy content of the flow, which experience shock increases as we increase the spin parameter of the BH.
In our low angular momentum, inviscid accretion flow model, there are two main driving forces, which act simultaneously on the flow. These forces are: inward attracting gravitational force and outward repulsive centrifugal force. 
With radial distance ($r$), the strength of these forces also varies.  If at a certain radius the outward centrifugal force dominates over the inward gravitational force then the supersonic flow will change its state of motion discontinuously and end up with the subsonic flow, that is the flow experiences a shock transition. Therefore, the centrifugal force manifests itself as the primary reason behind the shock transition in our model. Now, in the case of an axisymmetric rotating flow around a spinning BH, it is to be considered that the centrifugal force is a combined effect of the flow angular momentum  and spin angular momentum of the BH. Consequently, if we increase the spin parameter, the specific angular momentum of the flow should be reduced for admitting shock and the shock space moves towards the low angular momentum side. 

\begin{figure*}
    \centering
    \includegraphics[width=0.9\linewidth]{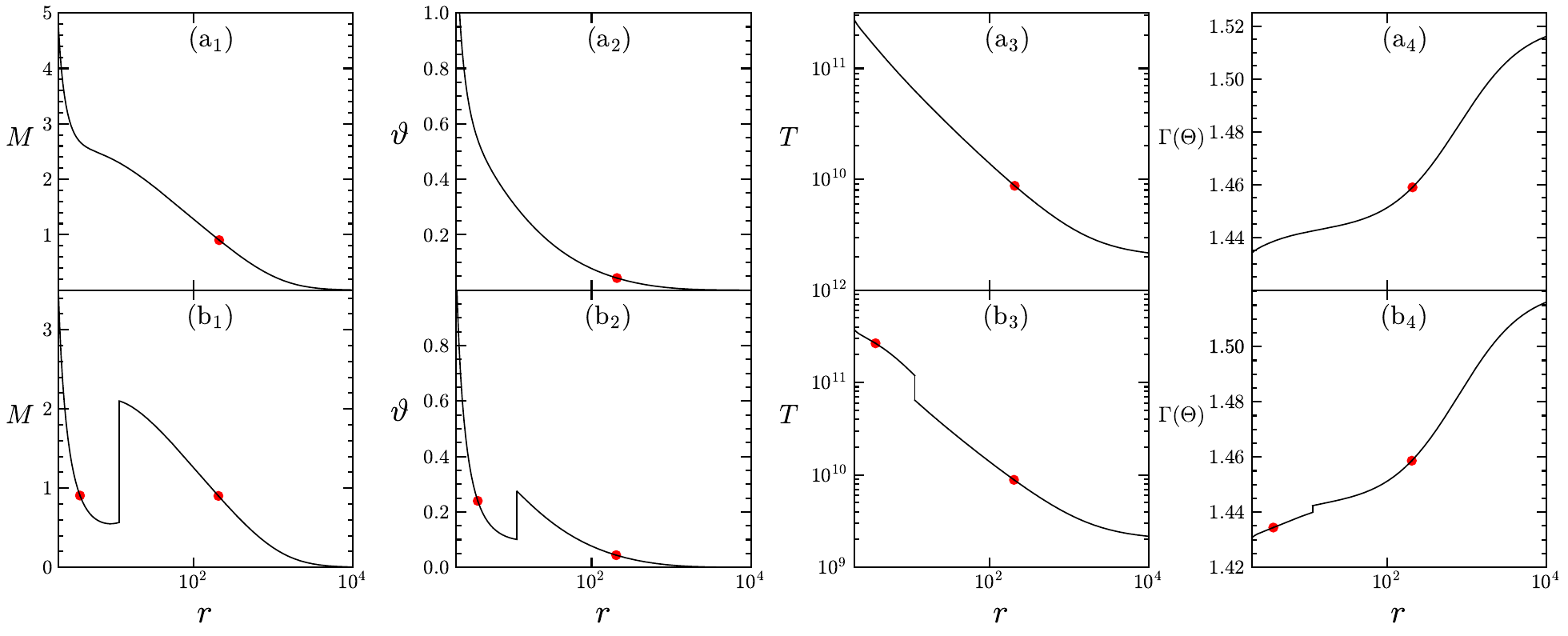}
    \caption{To clearly illustrate how the dynamical and thermodynamical quantities vary throughout the flow, we present in the upper panel: (a$_1$) Mach number $(M)$, (a$_2$) Radial velocity $(\vel)$, (a$_3$) Temperature $(T)$ and (a$_4$) Adiabatic index $(\Gamma(\Theta))$ as functions of the radial coordinate ($r$) for $\lambda = 2.0$. Similarly, in the lower panel, we show the variations of (b$_1$) Mach number $(M)$, (b$_2$) Radial velocity $(\vel)$, (b$_3$) Temperature $(T)$ and (b$_4$) Adiabatic index $(\Gamma(\Theta))$ with radial coordinate ($r$) for $\lambda = 2.33$. In all cases, we have chosen $\mathcal{E} = 1.001$ and $a = 0.6$. The solid red circle represents the location of the critical point. }
    \label{fig:7}
\end{figure*}

Fig.~\ref{fig:6}, describes different types of phase portraits ($M$ vs $r$ plot) that emerge from the dynamical system of axisymmetric inviscid rotating fluid flow around a spinning BH. Figs.~(6a$_1$ - 6a$_4$) show the phase diagram for BH spin parameter value $a = 0.0$ and specific angular momentum  $\lambda $ = 2.6 (Fig.~\ref{fig:6}a$_1$), 3.0 (Fig.~\ref{fig:6}a$_2$), 3.25 (Fig.~\ref{fig:6}a$_3$) and 3.6 (Fig.~\ref{fig:6}a$_4$). In each case, the solid black line depicts the global accretion solution. In Fig.~\ref{fig:6}a$_1$, the single saddle type critical point, far away from the horizon, is the outer critical point ($r_{\text{out}}$) and the flow passes only through this outer critical point and outer sonic point before crossing the BH horizon, which characterize it as a mono-critical, mono-transonic flow. In Fig.~\ref{fig:6}a$_2$, we have increased the angular momentum of the flow in such a way that an extra pair of critical points, one with saddle type (inner critical point, $r_\text{in}$) and other with centre-type (middle critical point, $r_\text{mid}$) in nature will form. 
In this case, the flow passes only through the outer critical and outer sonic point, so it is a multi-critical but monotransonic flow. In Fig.~\ref{fig:6}a$_3$, we increase the angular momentum further so that the flow admits shock due to the rise of the centrifugal barrier strength, and the supersonic flow will become subsonic again at the shock transition point. As a consequence of the shock transition, the flow becomes multi-critical and multi-transonic because it could now pass through both the outer and inner sonic point. Fig.~\ref{fig:6}a$_4$ shows that, further enhancement in angular momentum may cause the repulsive centrifugal force so strong that, it will take long distance for the fluid, from the outer boundary, to encounter its first critical point and evidently the  critical point will be very close to the horizon ($r_\text{in}$). Subsequently, the fluid could pass only through the inner critical point, and before encountering any shock it will reach the horizon, which makes it a monotransonic flow again. Figs.~(6b$_1$ - 6b$_4$) describe the phase portrait for BH spin parameter value $a = 05$ along with specific angular momentum $\lambda $ = 2.3 (Fig.~\ref{fig:6}b$_1$), 2.4 (Fig.~\ref{fig:6}b$_2$), 2.6 (Fig.~\ref{fig:6}b$_3$) and 2.9 (Fig.~\ref{fig:6}b$_4$). In Figs.~(6c$_1$ - 6c$_4$) we have plotted the phase diagram with BH spin parameter $a = 0.95$ and for specific angular momentum $\lambda $ = 1.3 (Fig.~\ref{fig:6}c$_1$), 1.5 (Fig.~\ref{fig:6}c$_2$), 1.6 (Fig.~\ref{fig:6}c$_3$) and 1.81 (Fig.~\ref{fig:6}c$_4$).

Next we explore how does the Mach number ($M$), radial velocity ($\vel$), temperature ($ T$) and adiabatic index ($\Gamma$) vary with radial distance ($r$) for two typical flow profile. For this discussion we have considered a different BH spin parameter value $a$ = 0.6 and specific energy $\mathcal{E}$ = 1.001. In the upper panel we have shown the variation of Mach number ($M$) (Fig.~\ref{fig:7}a$_1$), radial velocity ($\vel$) (Fig.~\ref{fig:7}a$_2$), temperature ($T$) (Fig.~\ref{fig:7}a$_3$) and adiabatic index ($\Gamma$) (Fig.~\ref{fig:7}a$_4$) with radial distance ($r$) for specific angular momentum parameter $\lambda$ = 2.0. The solid red circle denotes the location of the critical point, and the flow passes through only a single critical point as well as a single sonic point, making the flow a mono-transonic one. There is no discontinuity in any flow variables as we could see from the figures. In the lower panel we have plotted the change of Mach number ($M$) (Fig.~\ref{fig:7}b$_1$), radial velocity ($\vel$) (Fig.~\ref{fig:7}b$_2$), temperature ($T$) (Fig.~\ref{fig:7}b$_3$) and adiabatic index ($\Gamma$) (Fig.~\ref{fig:7}b$_4$) with radial distance ($r$) for specific angular momentum parameter $\lambda$ = 2.33. For the given parameter values, now the flow experience shock and is characterized by a multi-transonic flow. One crucial point to be noted from the figure, is the variation of the adiabatic index ($\Gamma$). It is also important to note that, in case of shocked flow the adiabatic index changes discontinuously at the shock location ($r_\text{sh}$).


\begin{figure*}
    \centering
    \includegraphics[width=0.9\linewidth]{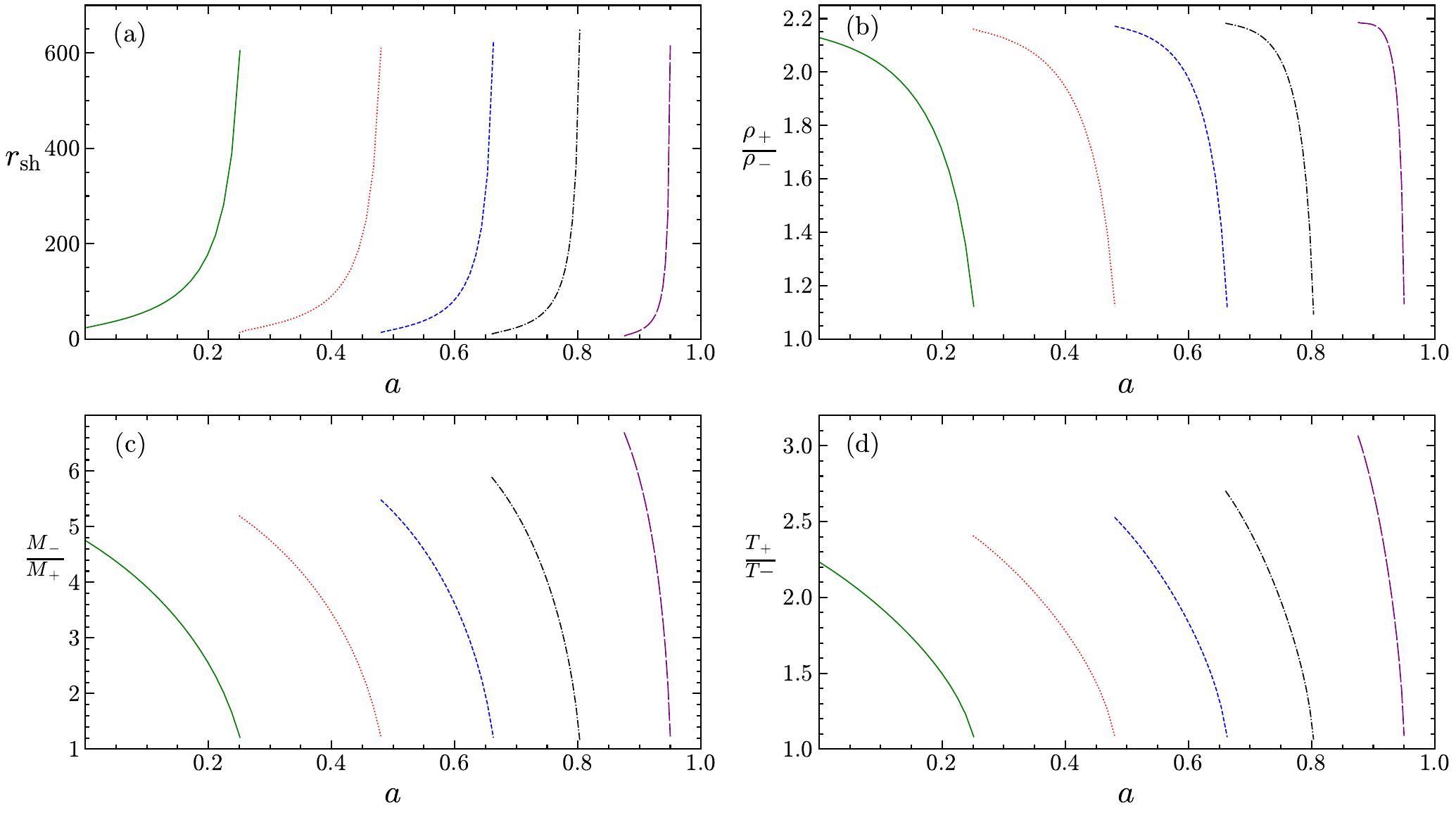}
    \caption{ Panel (a) shows the variation of shock location ($r_\text{sh}$)  with the BH spin parameter ($a$). Panel (b) presents the change in compression ratio ($\rho_{+}/{\rho_{-}}$) as a function of BH spin parameter ($a$). Panel (c) depicts the variation of shock strength ($M_-/M_+$)  with spin parameter ($a$), while panel (d) illustrates the variation of post-shock to pre-shock temperature ratio ($T_{+}/T_{-}$) with spin parameter ($a$). Each diagram corresponds to different specific angular momentum values: $\lambda$= 3.15 (solid green), 2.85 (dotted red), 2.56 (dashed blue), 2.275 (dashed dotted black) and 1.8 (long dashed purple). In all cases, we set $\mathcal{E} = 1.0001$ and $\xi = 1.0$.}
    \label{fig:8}
\end{figure*}

\begin{figure*}
    \centering
    \includegraphics[width=0.75\linewidth]{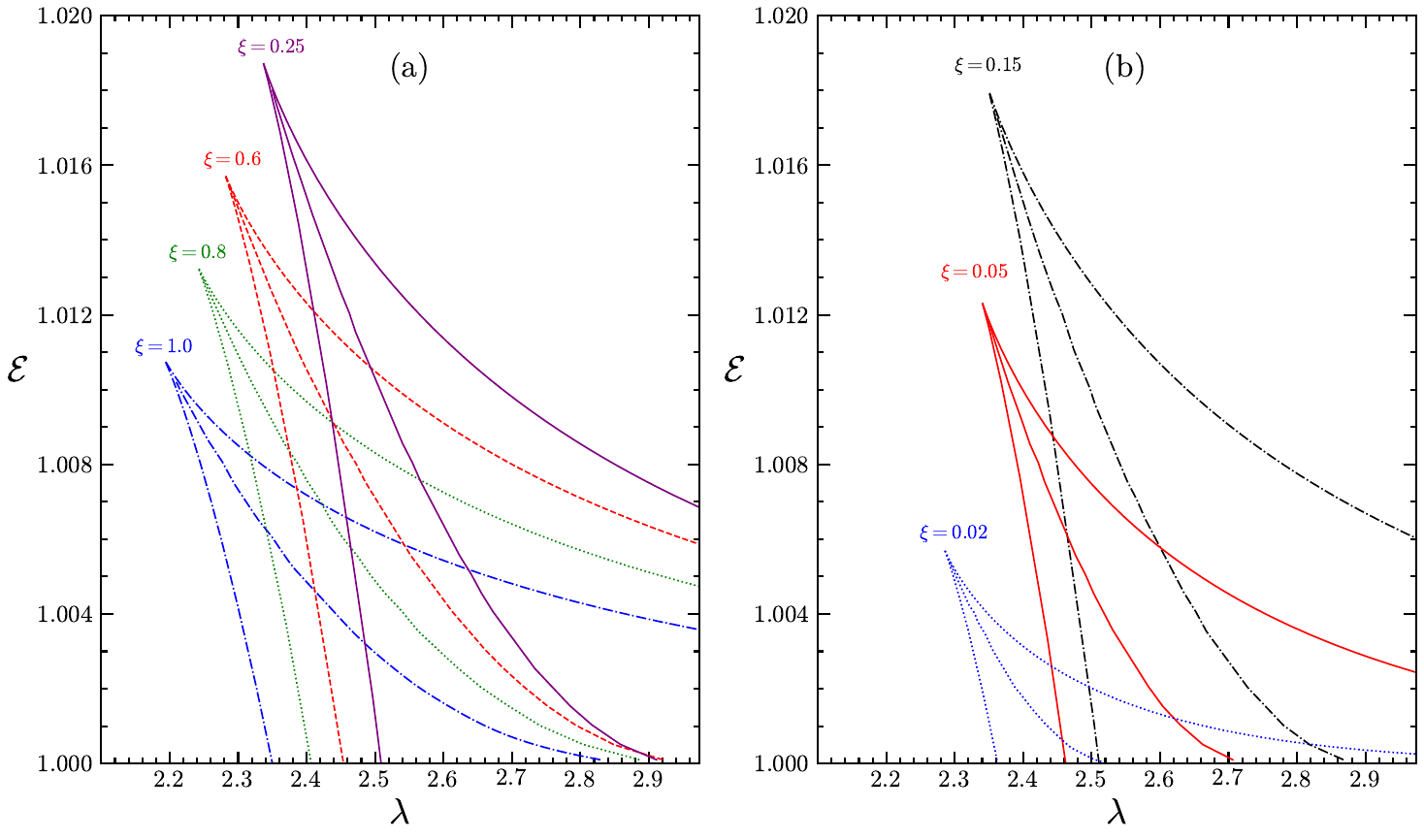}
    \caption{We have plotted the multi-critical parameter space for different values of composition parameter ($\xi$) in the ($\mathcal{E} - \lambda$) plane. In panel (a) $\xi = 1.0$ (dash-dotted blue), 0.8 (dotted green), 0.6 (dashed red) and 0.25 (solid purple), while in panel (b) $\xi = 0.15$ (dash-dotted black), 0.05 (solid red), 0.02 (dotted blue) and for $\xi = 0.0$ there is no multi-critical parameter space. We have plotted all the figures for $a = 0.5$.}
    \label{fig:9}
\end{figure*}


\begin{figure*}
    \centering
    \includegraphics[width=0.9\linewidth]{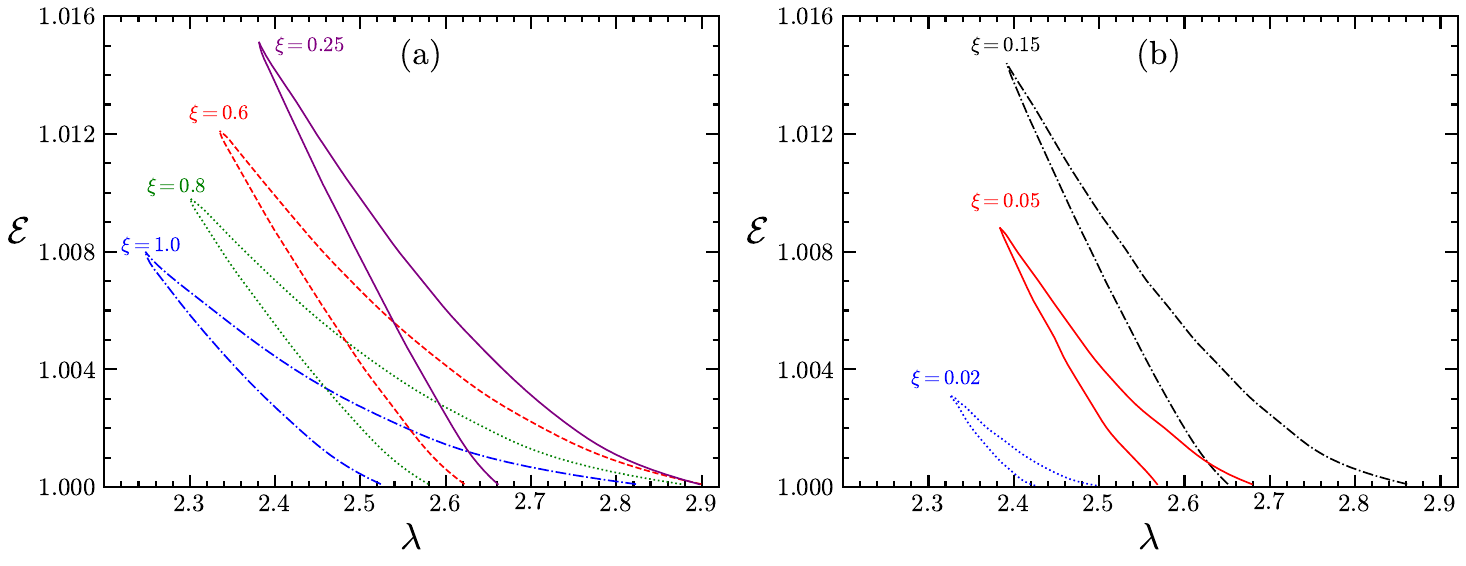}
    \caption{ Shock parameter space is plotted for different values of composition parameter ($\xi$) in the ($\mathcal{E} - \lambda$) plane. In panel (a) $\xi = 1.0$ (dash-dotted blue), 0.8 (dotted green), 0.6 (dashed red) and 0.25 (solid purple), while in panel (b) $\xi = 0.15$ (dash-dotted black), 0.05 (solid red), 0.02 (dotted blue) and for $\xi = 0.0$ we have not found any shock parameter space. Spin parameter value is fixed at $a = 0.5$ for all the figures.}
    \label{fig:10}
\end{figure*}

\subsection{Influence of spin on shock location and shock-induced flow variables}

The shock, which is a discontinuous and sudden change of the thermodynamical and dynamical variables of a flow, is of great importance in the astrophysical context. In the case of investigating low angular momentum, multi-transonic flow around a rotating BH, formation of shock  is an elemental feature of the fluid system. Various parameters, related to space-time geometry or the fluid itself, have colossal control over formation and orientation of shock. Interesting phenomena, related to BH accretion physics could be explained invoking the shock formation. Powering high-energy astrophysical jets, quasi periodic oscillation are some of these phenomena, which could be elucidated through shock formation. In this part, we probe the consequence of the change of BH spin parameter ($a$, which is linked to the space-time geometry) on shock location ($r_\text{sh}$) and various quantities of interest associated with shock.

In Fig.~\ref{fig:8}, for all the plots we have used composition parameter value $\xi$ = 1.0. In Fig.~\ref{fig:8}a, the variation of shock location ($r_\text{sh}$) has been plotted with the BH spin parameter ($a$) for fixed specific energy value $\mathcal{E} = 1.0001$ and specific angular momentum $\lambda$ = 3.15 (solid green), 2.85 (dotted red), 2.56 (dashed blue), 2.275 (dash-dotted black) and 1.8 (long dashed purple). Two prominent features could be observed from the figure. First, it is easy to notice from the diagram, that the shock location varies non-linearly with the spin parameter ($a$) and if the spin of the BH increases the shock forms at some larger distance from the BH horizon. Since, the shock forms due to the combined effect of inward attracting gravitational force and outward repulsive centrifugal force, if we increase the value of spin parameter for a fixed specific angular momentum ($\lambda$) of the flow, the outward centrifugal force will become stronger at each radius. As a result, the centrifugal force becomes comparable to the gravitational force at a larger radial distance. So, now it could stop the flow, which is nothing but the shock, at a larger radial distance from the horizon; therefore, the shock location ($r_\text{sh}$) shifts further away from the horizon. Second, after a certain limit of the spin parameter, if we increase its value for fixed specific angular momentum ($\lambda$), the stronger outward force will resist the inward moving fluid vigorously for the most part of its flow length. As a consequence the flow will pass only through the inner critical point ($r_\text{in}$) and shock won't form. Therefore, to cover the whole range of spin value, as we increase the spin parameter value, at the same time, we have to choose the lower specific angular momentum ($\lambda$) value accordingly. 

Now in Fig.~\ref{fig:8}b, we have explored the dependence of compression ratio, defined as the post-shock to pre-shock density ratio ($\rho_+/\rho_-$) with spin parameter for the aforementioned set of specific angular momentum values. The compression ratio varies non-linearly and inversely with the BH spin. For lower value of spin parameter with fixed angular momentum of the flow, the shock forms closer to the BH, as a result, the fluid gets more compressed and the value of compression ratio increases.

Next, we have plotted the variation of shock strength, defined as the ratio of pre-shock to post-shock Mach number ($M_-/M_+$) in Fig.~\ref{fig:8}c and post-shock to pre-shock temperature ratio ($T_+/T_-$) at the shock location in Fig.~\ref{fig:8}d, with BH spin parameter ($a$) for the same set of  specific angular momentum values as mentioned above. An obvious impression is that the shock strength ($M_-/M_+$) will be stronger for shock location ($r_\text{sh}$) closer to the BH horizon and that could be described from Fig.~\ref{fig:8}c unambiguously. The value of pre-shock Mach number ($M_-$) will be higher at the shock location situated very close to the horizon than that of a relatively large distanced shock location. Also, the energy released by the gravitational potential will be elevated at shorter shock locations, consequently, accessible energy for the bulk and thermal motion of the fluid also increases. In addition to that, due to shock the bulk velocity will be decreased drastically and the thermal energy part will increase. Therefore, the fluid gets hotter, making a rise in the post-shock to pre-shock temperature ratio ($T_+/T_-$) and the sound speed increases, which makes the Mach number lower and helps to increase the shock strength. So, in conclusion, we could assert that at a shorter shock location the fluid becomes hotter and more compressed and the shock strength and temperature ratio is anti-correlative with the BH spin parameter.

\begin{figure*}
    \centering
    \includegraphics[width=0.9\linewidth]{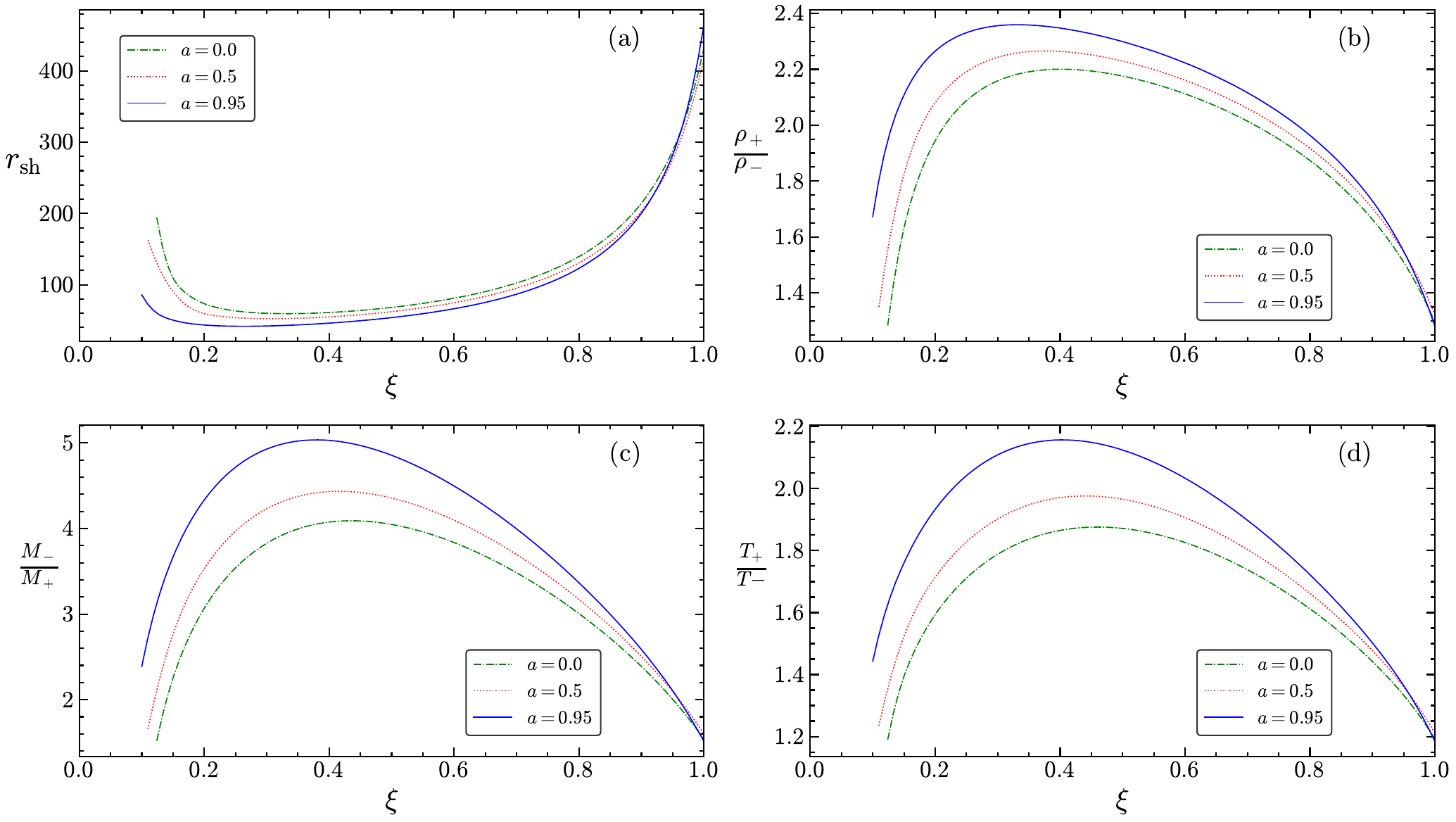}
    \caption{Panel (a) shows the variation of shock location ($r_\text{sh}$)  with the flow composition parameter ($\xi$). Panel (b) presents the change in compression ratio ($\rho_{+}/{\rho_{-}}$) as a function $\xi$. Panel (c) depicts the variation of shock strength ($M_-/M_+$)  with $\xi$, while panel (d) illustrates the variation of post-shock to pre-shock temperature ratio ($T_{+}/T_{-}$) with $\xi$. Each diagram corresponds to a different set of parameter values: [$a$, $\lambda$] = [0.0, 3.42] (dash-dotted green), [$a$, $\lambda$] = [0.5, 2.81] (dotted red) and [$a$, $\lambda$] = [0.95, 1.795] (solid blue). In all cases, we set $\mathcal{E} = 1.0001$.}
    \label{fig:11}
\end{figure*}



\subsection{Effect of fluid composition on the flow}

In this final section, we explore the effect of variation of the composition parameter ($\xi$) on some characteristic features of the flow. The admitting of multi-critical points and shocks are some typical features of the dynamics of low angular momentum axisymmetric fluid flow around a rotating BH. Therefore, we study the effect of the composition parameter ($\xi$) on multi-critical parameter space and on shock parameter space in explicit detail. Additionally, we talk about the dependence of shock location and some related quantities on the composition parameter.

In Fig.~\ref{fig:9}, we plot the multi-critical parameter space in the ($\mathcal{E}$ - $\lambda$) plane for different composition parameters. In Fig.~\ref{fig:9}a, we draw the multi-critical parameter space  for $\xi$ = 1.0 (dash-dotted blue), 0.8 (dotted green), 0.6 (dashed red) and 0.25 (solid purple), while in Fig.~ 9b, we plot it for $\xi = 0.15$ (dash-dotted black), 0.05 (solid red), 0.02 (dotted blue) and 0.0. In both of the cases, the spin parameter value is set to $a$ = 0.5.  We observe from the diagram that if we vary the composition parameter from $\xi$ = 1.0 to 0.25 (see Fig.~\ref{fig:9}a), the multi-critical parameter spaces get shifted in the higher specific angular momentum and specific energy region. Now, if we reduce the value of the composition parameter further from $\xi$ = 0.25 to 0.0 (see Fig.~\ref{fig:9}b), the multi-critical parameter spaces get shifted in lower specific angular momentum and specific energy region. For $\xi$ = 0.0, the flow is a pure $e^- - e^+$ flow (or pair plasma flow) and this type of flow does not show any dynamical nature containing multi-critical points.

Next in Fig.~\ref{fig:10}, we show the nature of the shock parameter space for the aforementioned set of composition parameter ($\xi$). For all the figures we fixed the BH spin parameter value at $a$ = 0.5. We notice similar type of behaviour for the shock parameter space with that of the multi-critical parameter space. If we decrease the value of the composition parameter from $\xi$ = 1.0 to 0.25 (see Fig.~\ref{fig:10}a), the shock parameter spaces move towards the higher specific angular momentum and specific energy region. If we reduce the value from $\xi$ = 0.25 to 0.0 (see Fig.~\ref{fig:10}b), the shock space now get shifted in the lower specific angular momentum and specific energy domain. For $\xi$ = 0.0, shock parameter space disappears, $\ie$, accretion flow with $\xi$ = 0.0, does not go through a shock transition.

Now, we discuss the effect of the composition parameter ($\xi$) on shock location ($r_\text{sh}$), compression ratio ($\rho_+/\rho_-$), shock strength($M_-/M_+$) and on post-shock to pre-shock temperature ratio ($T_+/T_-$). In Fig.~\ref{fig:11}a, we plot shock location with the composition parameter for the set of parameter values: [$a$, $\lambda$] = [0.0, 3.42] (dash-dotted green), [$a$, $\lambda$] = [0.5, 2.81] (dotted red) and [$a$, $\lambda$] = [0.95, 1.795] (solid blue). In Fig.~\ref{fig:11}b, we show the variation of compression ratio with $\xi$. Wile in Fig.~\ref{fig:11}c, we draw the variation of shock strength with $\xi$. Finally, we plot post-shock to pre-shock temperature ratio with $\xi$. For the last three figures, we use the previously mentioned parameter set for [$a$, $\lambda$]. In all the diagrams we set $\mathcal{E}$ = 1.0001.



\section{Concluding remarks}
In this work we have studied the basic features of the shocked accretion flow onto rotating BHs. Accreting matter has been described by a particular type of equation of state, and the steady state flow is assumed to take place under the influence of a particular type of post-Newtonian BH potential. Integral flow solutions have been constructed for axially symmetric flow maintained in hydrostatic equilibrium along the vertical direction.

Further extension of the present work is possible, along various different avenues, to obtain a more general scenario. Apart from flows in vertical equilibrium, axially symmetric flow of hydrodynamic matter can have various different geometrical configurations. Among those shapes, a disk-like flow with constant flow thickness, and a conical-shaped flow, are well studied in the literature \citep{2001MNRAS.327..808C, 2012NewA...17..285N, Saha_2016}. These three geometric configurations have been introduced using a set of idealised assumptions, though. A more realistic flow thickness is difficult to derive anyway, since such a task may be accomplished by employing the theory of non-LTE radiative transfer or by incorporating the Grad-Shafronov equations for complete MHD/plasma flows \citep{1997PhyU...40..659B, 2009mfca.book.....B, Beskin_2005, 1998ApJ...505..558H, 2006ApJS..164..530D}. 

In our present work, we have not performed any stability analysis of the transonic, stationary integral flow solutions, and hence, we are not in a position to make comments whether the steady state solutions obtained in the present work are compatible with observed astrophysical events with duration of some specified time scale.

Our immediate next project will thus be a more comprehensive study of the shocked accretion solutions for all available geometrical configurations of the axially symmetric flow, and to investigate their linear stability properties as well, to understand how such stability criteria, as well as the emergent acoustic metric, can be influenced by the spin angular momentum of the BHs, $\ie$, the Kerr parameter $a$.

It is to be noted that the shocked accretion solutions for steady state accretion as described by Ryu and his collaborators can be studied for full general relativistic flow as well \citep{ck16,kc17,scp20,sc22}, we still perform our investigation of such flows for post-Newtonian BH potentials. Although the stationary integral solutions can be constructed for general relativistic Kerr space-time, it is surely not assumed that the complete flow profile with comprehensive spectral details can be obtained within complete general relativistic framework, rather a pseudo-Kerr formalism helps to formulate and solve the corresponding flow equations when one plans to take a step further by studying the radiative and spectral properties of the flow in addition to its dynamics only. We thus believe that it is imperative to introduce a comprehensive formalism capable to study the accreting BH systems within the framework of post-Newtonian BH potentials. In the present work, we have performed our calculation for the BH potential as introduced by Artemova et. al \citep{1996ApJ...461..565A}, since by far this one is the simplest looking pseudo-Kerr potential available in the literature. However, several other pseudo-Kerr potentials have been introduced by various authors \citep{1992MNRAS.256..300C, lovaas1998modified, 1999A&A...343..325S, Mukhopadhyay_2002, 10.1111/j.1365-2966.2006.10350.x, ghosh2007generalized, Ghosh_2014, 2014bhns.work..121K}. We plan to study the shocked accretion flow and to perform its linear stability analysis for all such pseudo-Kerr potentials in future.











\appendix

\section{Mass conservation equation}\label{def_Sigma}
In general mass conservation equation for a fluid system is given by the following form:
\begin{equation}
    \frac{\partial \rho}{\partial t} + \nabla .(\rho \boldsymbol{\vartheta}) = 0
    \label{general_mass_con.eq}
\end{equation}
For our study we choose cylindrical polar coordinate system ($r$, $\phi$, $z$), in which equation (\ref{general_mass_con.eq}) takes the below form:
\begin{equation}
   \frac{\partial \rho}{\partial t} + \frac{1}{r}\frac{\partial}{\partial r}(r\rho \vel_\mathrm{r}) + \frac{1}{r}\frac{\partial}{\partial \phi}(\rho \vel_\mathrm{\phi}) + \frac{\partial}{\partial z}(\rho \vel_\mathrm{z}) = 0
   \label{cylindrical_mass_con.eq}
\end{equation}
where $\vel_\mathrm{r}$ is the radial component, $\vel_\phi$ is the azimuthal component and $\vel_\mathrm{z}$ is the $z$ component  of the velocity vector ($\boldsymbol{\vel}$). For convenience, we write the radial component of velocity ($\vel_\mathrm{r}$) as simply $\vel$. 

As our system is  axially symmetric along the $z$ axis, variables are independent of the $\phi$ coordinate. So, the third term of equation~(\ref{cylindrical_mass_con.eq}) vanishes. We also assume that our flow structure is in hydrostatic equilibrium in the $z$ direction and the value of $\vel_\mathrm{z}$ is negligible compared to the other velocity components. Therefore, we set $\vel_\mathrm{z} = 0$, which causes the last term in equation (\ref{cylindrical_mass_con.eq}) to vanish. Under these assumptions, equation (\ref{general_mass_con.eq}) could be rewritten as follows:
\begin{equation}
    \frac{\partial \rho}{\partial t} + \frac{1}{r}\frac{\partial}{\partial r}(r\rho \vel) = 0
    \label{new_mass_con.eq}
\end{equation}

Now, it is a common practice in accretion flow study to vertically integrate (in our case it is along the $z$ direction) the governing equation/fluid variables to remove any vertical dependency (see \citealt{2014pafd.book.....C}, \citealt{1998ApJ...498..313G}). Therefore, we integrate equation (\ref{new_mass_con.eq}) along the $z$ direction and recast it in the following form:
\begin{equation}
   \frac{\partial \Sigma}{\partial t} + \frac{1}{r}\frac{\partial (\Sigma \vel r)}{\partial r} = 0 
\end{equation}
where, $\Sigma = \int^{+H}_{-H} \rho \,dz = 2\rho H$, is the vertically averaged surface density over the disk height and $\rho$ is fluid density at the equatorial plane ($\ie$, $z = 0$ plane). Here, $H$ is the disk half height (see Appendix \ref{disk_height}).

\section{Disk height calculation}\label{disk_height}
For axisymmetric inviscid steady fluid flow, in cylindrical polar coordinate system we have to solve the radial component ($r$ - component) and the vertical component ($z$ - component) of the Euler equation to generate the flow structure. The radial component of the Euler equation is given in equation (\ref{eu_steady.eq}) and we  solve this on the equatorial plane and now we write down the $z$ component of Euler equation as follows:
\begin{equation}
     \vel \frac{\partial \vel_\mathrm{z}}{\partial r} + \vel_\mathrm{z} \frac{\partial \vel_\mathrm{z}}{\partial z} + \frac{1}{\rho}\frac{\der p}{\der z} + \frac{\partial \Phi_\mathrm{ABN}}{\partial z} = 0
     \label{z_euler.eq}
\end{equation}

One important thing to be noted is that the form of the pseudo-potential (equation \ref{ABN_potential.eq}) that we have taken to solve the radial component of Euler equation is defined on the equatorial plane, $\ie$, $z = 0$ plane. Now, to solve the Euler equation in the vertical direction, we have to take the general form of the pseudo-potential in cylindrical polar coordinate system. The general form of the potential in cylindrical polar coordinate system ($r$, $\phi$, $z$) is given as:
\begin{equation}
   \Phi_\text{ABN}=\frac{1}{(1-\beta) \rp}\left(1-\frac{\rp}{\sqrt{r^2 + z^2}}\right)^{1-\beta}-\frac{1}{(1-\beta) \rp}
   \label{general_potential.eq}
\end{equation}

Now, it is very challenging if not impossible to analytically solve both the radial and vertical components of the Euler equation simultaneously to construct the flow structure. Consequently, people usually prefer to approximate the vertical flow structure using some kind of height function (for discussion, see \citealt{Nag} and the references therein) rather than solving the vertical component of the Euler equation. Here, we will construct the height function by approximating the vertical Euler equation by claiming that the flow is in hydrostatic equilibrium in the vertical direction. Under this consideration we could write equation (\ref{z_euler.eq}) in the following way:
\begin{equation}
    \frac{1}{\rho}\frac{\der p}{\der z} + \frac{\partial \Phi_\mathrm{ABN}}{\partial z} = 0
    \label{hydrostatic_equilibrium.eq}
\end{equation}
If we take derivative of $\Phi_\mathrm{ABN}$, given in equation (\ref{general_potential.eq}), with respect to $z$, along with the criteria that $z\ll r$, then we get:
\begin{equation}
    \frac{\partial \Phi_\mathrm{ABN}}{\partial z} = -\frac{zF_\mathrm{ABN}}{r}
    \label{dphi_dz.eq}
\end{equation}
See equation (\ref{f_ABN.eq}) for the expression of $F_\mathrm{ABN}$. Here, we consider that $p$ will vary linearly in the vertical direction, $\ie$, in the $z$ direction and we approximate ${\der p}/{\der z}$ as:
\begin{equation}
    \frac{\der p}{\der z} = -\frac{p}{z}
    \label{dp_dz.eq}
\end{equation}
Where, the `--' sign takes care of the fact that pressure ($p$) decreases away from the equatorial plane ($\ie$, $z = 0$ plane). Using equation (\ref{dphi_dz.eq}) and equation (\ref{dp_dz.eq}) in equation (\ref{hydrostatic_equilibrium.eq}) we get:
\begin{equation}
    z^2 = -\frac{p}{\rho}\frac{r}{F_\mathrm{ABN}}
\end{equation}
After some manipulation, the expression for $z$ is given as:
\begin{equation}
    z = c_\mathrm{s}\sqrt{\frac{ r}{\Gamma |F_\mathrm{ABN}|}}
    \label{z_height.eq}
\end{equation}
In equation (\ref{z_height.eq}), $z$ is nothing but the height ($H$) of the flow.

\section{Various coefficients}\label{coefficients}

Here we write down the functional forms of the coefficients that we have used in equations (\ref{dN_dr}), (\ref{dD_dr}) and (\ref{root_equation}):\\

$\begin{aligned}
\mathcal{N}_{11} = & -\frac{2\Theta\Gamma}{\tau(\Gamma + 1)}\bigg[  \frac{(5 - \beta)}{r^2} + \frac{\beta}{(r -\rp)^2}\bigg] - \frac{3\lambda^2}{r^4}\\ 
&+ \frac{2r -(2 - \beta)\rp}{r^{3 - \beta}(r - \rp)^{\beta + 1}} \\
\mathcal{N}_{12} = &  \frac{2}{\tau(\Gamma + 1)}\bigg[ \frac{(5 - \beta)}{r} + \frac{\beta}{(r - \rp)}        \bigg]\bigg[ \Gamma + \frac{\Theta}{(\Gamma + 1)}\frac{\der \Gamma}{\der \Theta}   \bigg]       \\ 
\mathcal{D}_{11} = & 1 + \frac{4\Theta\Gamma}{\tau(\Gamma + 1)\vel^2}\\
\mathcal{D}_{12} = & -\frac{4}{\tau(\Gamma + 1)\vel}\bigg[ \Gamma + \frac{\Theta}{(\Gamma + 1)}\frac{\der \Gamma}{\der \Theta}  \bigg]
\end{aligned}$\\

where, all quantities retain their usual definitions.

\section{shock invariant quantity calculation}\label{shock_invariant}
Here we explain how to compute the shock invariant quantity ($I$) using the Rankine -- Hugoniot shock conditions, $\ie$, equations (\ref{m_shock.eq}) -- (\ref{energy_shock.eq}). Equation (\ref{m_shock.eq}) could be written as:
\begin{equation}
    \rho_{+}\vel_{+}H_{+} = \rho_{-}\vel_{-}H_{-}
    \label{m_shock_new.eq}
\end{equation}
whereas explicit form of equation (\ref{mom_shock.eq}) is:
\begin{equation}
   H_{+}p_{+} + H_{+}\rho_{+}\vel_{+}^2 = H_{-}p_{-} + H_{-}\rho_{-}\vel_{-}^2 
   \label{mom_shock_new.eq}
\end{equation}
Using equation (\ref{density.eq}) and equation (\ref{pressure.eq}) one could write:
\begin{equation}
    p = \frac{2\Theta\rho}{\tau}
    \label{p_density_relation.eq}
\end{equation}
Plugging equation (\ref{p_density_relation.eq}) into equation (\ref{mom_shock_new.eq}), we get:
\begin{equation}
    \rho_{+}\vel_{+}H_{+}\bigg[ \frac{2\Theta_{+}}{\tau\vel_{+}} + \vel_{+}    \bigg] = \rho_{-}\vel_{-}H_{-}\bigg[ \frac{2\Theta_{-}}{\tau\vel_{-}} + \vel_{-}    \bigg]
\end{equation}
Now, by taking the help of equation (\ref{m_shock_new.eq}) and after some manipulation we come up with the following condition:
\begin{equation}
    \frac{1}{\vel_{+}}\bigg[ \Theta_{+} + \frac{\tau}{2}\vel_{+}^2  \bigg] = \frac{1}{\vel_{-}}\bigg[ \Theta_{-} + \frac{\tau}{2}\vel_{-}^2  \bigg]
\end{equation}
The above condition, known as the shock location condition, helps us to find the location of shock ($r_\mathrm{sh}$), whereas the quantity defined by:
\begin{equation}
    I = \frac{1}{\vel} \bigg[ \Theta + \frac{\tau}{2}\vel^2  \bigg]
\end{equation}
is known to be the shock-invariant quantity. The numerical value of the shock invariant quantity will be the same only at the location of shock. Therefore, using the quantity $I$, one could easily find the shock location for multi-transonic accretion flow.


\bsp	
\label{lastpage}
\end{document}